\documentclass[aps,pra,twocolumn,10pt,nofootinbib,superscriptaddress]{revtex4-2}
\usepackage[T1]{fontenc}
\usepackage[utf8x]{inputenc}
\usepackage{lmodern}
\usepackage[english]{babel}
\usepackage{indentfirst}
\usepackage{amsmath}
\usepackage{amsfonts}
\usepackage[table]{xcolor}
\usepackage{framed}
\usepackage[framemethod=tikz]{mdframed}
\usepackage[framed,amsthm]{ntheorem}
\usepackage{dsfont}
\usepackage{cancel}
\usepackage{float}
\usepackage{graphicx}
\usepackage[normalem]{ulem}

\theoremstyle{remark}

\usepackage[colorlinks,linkcolor=blue,citecolor=blue,urlcolor=blue]{hyperref}

\newcommand{\id}{\ensuremath{\mathds{1}}}

\newcommand{\prob}{\operatorname{Prob}}

\newcommand{\bra}[1]{\langle #1|}
\newcommand{\ket}[1]{|#1\rangle}
\newcommand{\braket}[2]{\langle #1|#2\rangle}

\newcommand{\bea}{\begin{eqnarray}}
\newcommand{\eea}{\end{eqnarray}}

\newcommand{\bi}{\begin{itemize}}
\newcommand{\ei}{\end{itemize}}

\newcommand{\si}{\sigma}

\newcommand{\cE}{{\mathcal E}}

\newcommand{\CC}{{\mathbb C}}

\newcommand{\tr}{\operatorname{tr}}

\newcommand\errch{\mathcal{E}}
\newcommand\pauch{\mathcal{S}}

\newcommand{\CZ}{\mathsf{CZ}}
\newcommand{\FSWAP}{\mathsf{fSWAP}}
\newcommand{\SWAP}{\mathsf{SWAP}}

\newcommand{\berlinaffil}{Dahlem Center for Complex Quantum Systems, Freie Universit{\"a}t Berlin, 14195 Berlin, Germany}
\newcommand{\ibkaffil}{University of Innsbruck, Institute for Theoretical Physics, A-6020 Innsbruck, Austria}
\newcommand{\tumaffil}{Technical University of Munich, TUM School of Natural Sciences, Physics Department, 85748 Garching, Germany}

\begin{document}

\title{Gaining confidence on the correct realization of arbitrary quantum computations}

\author{Jose Carrasco}
\affiliation{\berlinaffil}
\affiliation{\ibkaffil}
\author{Marc Langer}
\affiliation{\tumaffil}
\affiliation{\ibkaffil}
\author{Antoine Neven}
\affiliation{\ibkaffil}
\author{Barbara Kraus}
\affiliation{\tumaffil}
\affiliation{\ibkaffil}

\date{\today}
\begin{abstract}We present verification protocols to gain confidence in the correct performance of a device implementing an arbitrary quantum computation. The derivation of the protocols is based on the fact that matchgate computations, which are classically efficiently simulable, become universal if supplemented with additional resources. We combine tools from weak simulation, randomized compiling, and statistics to derive verification circuits that (i) strongly resemble the original circuit and (ii) can be classically efficiently simulated not only in the ideal, i.e. error free scenario, but also in the realistic situation where errors are present. In fact, in one of the protocols we apply exactly the same circuit as in the original computation, however, to a slightly modified input state. 
\end{abstract}
\maketitle

With the advent of ever larger quantum processors, the
question of how to evaluate their performance becomes
increasingly relevant. A distinction is made between pro-
tocols in which the quantum device is potentially untrusted by the user, such as a cloud computer, from those
in which one does have direct access to the quantum processor. In
the former scenario a solution based on interactive proofs and post-quantum cryptography has been presented \cite{Mah, Vidick1, Vidick2}. In the latter,  several protocols to gain confidence in the performance of a quantum device have been put forward (for a recent review, see~\cite{Ei20}). Randomized benchmarking \cite{KnillRB, LuLi15} and several variants of it have been developed \cite{MaGa12, OnWe19, MirrorRB} where the average performance of e.g. Clifford gates is quantified with a single parameter, the average gate fidelity \cite{NC10}.

Also due to their importance in fault tolerant quantum computation, the simulability of Clifford circuits has been utilized to study the performance of particular quantum computations~\cite{JS17,FeKa19} and bound the total variation distance of the erroneous to the ideal output state~\cite{FeKa19}. However, in order to accomplish the challenging task of verifying universal quantum computations and to also take into account the gates that naturally occur in e.g. real-time Hamiltonian evolution, which might be difficult to benchmark with other methods~\cite{ScCi08}, different gate sets need to be considered. Moreover, the problem of characterizing the reliability of implementing a particular (not on average) universal computation with more than a single error parameter is largely unexplored. It is precisely this problem that we will address in the present work: To check not only the computation itself but to also gain confidence in the correctness of an entire error model.

We consider an arbitrary quantum computation and analyze how the distribution of single, or multiqubit measurement outcomes can be tested. One of the main obstacles here is, of course, that the correct outcome is unknown, as the computation can not be performed classically efficiently\footnote{Note that there are however methods to compare the output of two computations (quantum and/or classical), such as cross--platform verifications \cite{ElbenZoller}.}. Moreover, as it is (in the near future) inevitable that errors will occur during the computation, a classical simulation of the ideal  computation would only provide a limited amount of information on the quality of the quantum device. Only the simulation of the erroneous computation, which can be parameterized by a set of error parameters, allows not only to establish trust in the outcome, but also in the error model. A natural issue arising here is that even if the computation itself were efficiently simulable, inclusion of arbitrary -- possibly coherent -- errors likely renders the simulation hard.

\begin{figure*}[htb!]
\centering
\includegraphics[width=\linewidth]{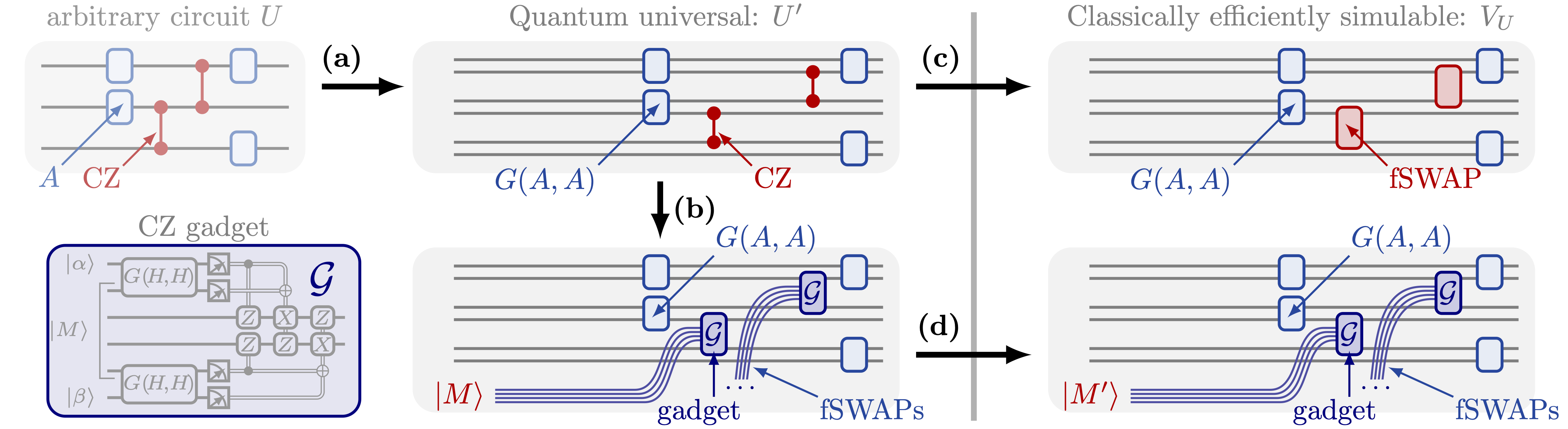}
\caption{Verification circuits:
(a) $U$ is mapped to its encoded version, $U'$, which is acting on $2m$ qubits and is decomposed into n.n. MGs and $\CZ$ gates; (b) each resourceful $\CZ$ gates can be implemented deterministically using the magic states $\ket M$ and adaptive measurements via the gadget (bottom left)  \cite{Hebenstreit2019}. To verify the encoded realizations of $U$ through efficient classical simulations, we consider verification circuits obtained either by (c) replacing all $\CZ$-gates with $\FSWAP$ gates but leaving the structure of the circuit intact, or (d) implementing exactly the same circuit but on a slightly modified input state.}
\label{fig:results}
\end{figure*}

We will show that combining methods from weak simulation of quantum computation \cite{Ne11} with randomized compiling \cite{EmersonRandComp} and classical statistics allows one to overcome these obstacles. For a given (universal) quantum computation we will introduce verification protocols, which test computations which differ only in some gates, or the input state, compared to the original computation (see Fig.~\ref{fig:results}). We will show that via the notion of randomized compiling the output state of the erroneous quantum computation can be tailored into one which is parameterized by a few error parameters and is, crucially, still weakly efficiently simulable. This holds under mild conditions on the error model. The circuits we verify must be classically efficiently simulable and therefore must, in general, differ from the original circuit. However, we choose our verification circuits very similar to the original circuit. In fact, we either only add some additional gates, or apply exactly the same computation on a slightly different input state. Then, we will show that tests from classical statistics allow us to estimate the error parameters of the (slightly modified) computation (see Fig.~\ref{fig:results}). Hence, as an output of these tests we obtain the error parameters which completely characterize the output state of the slightly modified (randomly compiled) circuit. This can then also be utilized to verify the error model, to detect e.g. possible drifts during the computation, and might also be used in error mitigation. We illustrate the performance of the verification protocols with some examples, where we also present some additional tests.

Central to our approach are matchgate circuits. A matchgate (MG) is a two--qubit gate which can be written as $G(U_1,U_2)=U_1\oplus U_2$. Here, $U_1$ ($U_2$) is a unitary acting on the even (odd) parity subspace respectively and their determinants coincide. Matchgate circuits (MGCs) are composed out of nearest neighbor (n.n.) MGs acting on computational basis states and the output corresponds to the measurement of a single (or several) qubit(s) in the computational basis. MGCs can be simulated classically efficiently \cite{Valiant, TerhalMG, JozsaMiyake} and even compressed into an exponentially smaller quantum computer \cite{JoszaMiyakeBKWotrous}. Which resources can be added to MGCs such that the computation remains classically efficiently simulable has been studied in \cite{Br16,Hebenstreit2020}. One distinguishes here between strong and weak simulation. Whereas strong simulation means that for any given output bitstring $z$ on any subset of measured qubits, $p(z)$ can be determined classically efficiently, weak simulation implies that one can classically sample form the exact output probability distribution \cite{Ne11} (see also Appendix~\ref{appendix:simulation-of-mgs}).

There are several ways to elevate the computational power of MGC to the one of a universal quantum computer. To review them, we consider here and throughout the paper an arbitrary, universal circuit, $U$, of width $m$ (number of qubits it is acting on) and size $poly(m)$. $U$ can be decomposed into single qubit and $r=\Omega(poly(m))$ controlled two--qubit phase gates, $\CZ$. Slightly modifying the encoding used in \cite{JozsaMiyake,Brod2011ExtendingMGs} it is easy to show that such a circuit can be encoded into a circuit $U'$ of width $n=2m$, which is composed out of $poly(m)$ n.n. MGs and $r$ resourceful n.n. $\CZ$ gates. This can be easily seen using the freely swappable logical states $\ket{00}$ and $\ket{11}$, the encoding of the single qubit gates $A_i$, $G(A,A)_{2i-1,2i}$, and $\CZ_{2i,2i+1}$ gates acting between logical states (see Appendix~\ref{appendix:mgc-measurements-and-magic-states}). Here and in the following, the subscript denotes the qubit(s) the gate is acting on. Thus, supplementing MGCs with n.n. $\CZ$ gates leads to universal quantum computation. 

The resourceful gate can also be deterministically implemented via gate teleportation using magic states and adaptive measurements. As shown in \cite{Hebenstreit2019}, any non--Gaussian fermionic state is a magic state for MG computations. i.e. is resourceful. For instance, the magic state, $\ket{M}=\CZ_{2,3} \ket{\Phi^+}_{1,2} \ket{\Phi^+}_{3,4}$ can be utilized together with adaptive measurements to implement the $\CZ$ gate deterministically\footnote{Let us stress here that using this magic state, the $\CZ$ gate can also be implemented on non-n.n. qubits (see Appendix~\ref{appendix:mgc-measurements-and-magic-states}).} \cite{Hebenstreit2019}. Crucial for our approach will be that supplementing MGCs with any of these ingredients separately, i.e. magic states with at most ${\cal O} (\log(n))$ adaptive measurements or adaptive measurements with at most ${\cal O} (\log(n))$ magic states, remains classically efficiently simulable \cite{Hebenstreit2020}.

Using the results summarized above we will now derive verification circuits, for which the erroneous output can be simulated classically efficiently \footnote{Note that, in \cite{JS17} a similar approach has been presented for Clifford gates and the measurement of a single qubit. However, there, the single qubit reduced states are completely mixed (or factorize). The here proposed method might, however, be applicable to Clifford circuits.}. Let us first explain two methods to map the fixed, but arbitrary  encoded circuit, $U'$, to a slightly modified classically simulable circuit $V_U$ (see Fig.~\ref{fig:results}). Starting with $U'$ and replacing each of the $\CZ$ gates by a $\FSWAP$ gate, i.e. by $\CZ\cdot\SWAP$, leads to a classically efficiently simulable circuit. Albeit this is a very simple mapping, it is clear that exchanging the $\CZ$ gates with $\FSWAP$ gates is a drastic change in the computation. A more sophisticated mapping in which the circuit is exactly the same, but the gates are applied to a slightly different input state is the following. As explained above, $U'$ can be realized via an adaptive circuit composed out of MGs applied to the input state $\ket{0}^{\otimes n}\ket{M}^{\otimes r}$. Considering exactly the same circuit, including adaptive measurements (using exactly the same correction operators, or modified ones to implement e.g. deterministically the $\FSWAP$ gate, see Appendix~\ref{appendix:mgc-measurements-and-magic-states}), but applied to $\ket{\Psi_\mathrm{in}}=\ket{0}^{\otimes n}\ket{M'}^{\otimes r}$  with $\ket{M'}=\CZ_{2,3} \ket{\Phi^+}_{1,3} \ket{\Phi^+}_{2,4}$ leads to a circuit which is classically efficiently simulable. The reason for that is that the state $\ket{M'}$ can be generated with MGs (in contrast to the state $\ket{M}$) and that adaptive measurements on those states remain classically efficiently simulable \cite{TerhalMG, Hebenstreit2020}.

We show next, how the circuit can be transformed into one which allows for the efficient simulation of the erroneous realisation of $V_U$. To this end, we employ the notion of Randomized Compiling (RC) \cite{EmersonRandComp}. RC does not only lead to a more robust implementation of the circuit, but, as we will show, allows us to tailor the output of the quantum computation to a state whose output probability distribution can be sampled from classically efficiently. That is, we show now that we can weakly simulate the output of the randomly compiled, erroneous realization of $V_U$. Using statistical tests, such as the Kolmogorov-Smirnov (KS) \cite{Ko33} or the Epps-Singleton (ES) \cite{EppsSingleton} test (see below) enables us to compare the samples and to gain confidence that they stem from the same probability distribution. Altogether, this allows us to estimate the error parameter(s) of the randomly compiled computation. 

We will make the following assumptions on the error model: (i) Instead of a MG, $M$, the map $\Lambda_M$ is implemented, where $\Lambda_M(\rho) = \errch_M(M \rho M^\dagger)$ and the error $\errch_M$ can depend on $M$, but is assumed to be Markovian; (ii) Pauli operators can be implemented with negligible error; (iii) a measurement with projectors $\Pi_t, t\in\{0,1\}$, is modelled by $\Pi_t \mathcal{E}(\cdot) \Pi_t$; (iv) for any MG, $M$ and any $k$--fold Pauli operator $P$ it holds that $\Lambda_{M(P)}={\errch}_M (M(P) \cdot  M(P)^\dagger) $, where $M(P)=PMP$. Here, ${\errch}_M$ depends on $M$, but is independent of $P$. Additionally, we assume that any error channel acts on at most at most ${\cal O}(\log(n))$ qubits and that the initial state $\ket{0^n}$ can be prepared perfectly.
The first three assumptions are not very stringent and are commonly used \cite{EmersonRandComp, MirrorCircuitsTrust}. To see that assumption (iv) is justified note that any MG is, up to local phase gates ($e^{i \alpha_i Z}$) of the form $e^{i\beta XX +i\gamma YY}$ \cite{Murao}. Thus, for any Pauli operator $P$, acting on arbitrary many qubits, we have $PM(P)P=M$, where the local and non-local parts of $M$ and $M(P)$ coincide up to changing the signs of the phases ($\alpha_i$,  $\beta$ and $\gamma$) randomly, which justifies assumption (iv). It follows that error models $\errch_M$ depending on the absolute value of the mentioned angles do satisfy assumption (iv). This encompasses coherent errors represented by over-rotations in the form of $e^{\pm i|\alpha_i|\epsilon Z}$, $e^{\pm i|\beta|\epsilon XX}$, or $e^{\pm i|\gamma|\epsilon YY}$. Additionally, it includes stochastic errors where each over-rotation occurs with a certain probability. Let us conclude this discussion on the error model by noting that our assumptions can be relaxed, such that the noisy state remains simulable. This can be achieved in two ways: Firstly, after randomized compilation, any Pauli channel would be admissible, provided the total number of parameters is at most $poly(n)$ (including some errors that are correlated in time). Secondly, one can allow for errors that are convex combinations of MGs, e.g. certain over-rotation errors, as this remains simulable.

Next, we show that under these assumptions on the error model, one can depolarize the error of any MGC to a Pauli channel. For each MG, $M_i$ we choose a random Pauli operator, $P_i \in \mathcal{P}_n$, and apply the gate sequence $P_iM_i(P_i)P_i$. In the error--free case we obtain the final pure state $\prod_{i=1}^sP_iM_i(P_i)P_i\ket{0\cdots 0}$ which coincides with the ideal state $\prod_{i=1}^s M_i\ket{0\cdots0}$ as $P_iM_i(P_i)P_i=M_i$ for each $i$. To analyze the erroneous case we consider first a single gate, $M_i$ with  corresponding error channel $\errch_i$. As shown in Appendix~\ref{appendix:randomized-compilation}, $\errch_i(\cdot)$ is transformed to a Pauli channel $\pauch_i$, i.e. $|\mathcal{P}_n|^{-1}\sum_{P_i} P_i\errch_M(P_i\cdot P_i) P_i=\pauch_i(\cdot)$. Concatenating the channels for the whole circuit leads to the output state
\begin{equation}
\begin{aligned}
&\rho_{\rm exp}=\sum_{\{P_i\}}c_1(P_1)\cdots c_s(P_s)W(P_1,\ldots,P_s)\,, \end{aligned}
\label{eq:maintext_rho_out}
\end{equation}
where $W(P_1,\ldots,P_s)$
denotes the projector onto the state
\begin{equation}
  \begin{aligned}
    \ket{\psi(P_1,\ldots,P_s)}=P_sM_s\cdots P_2M_2P_1M_1&\ket{0^n}\\
    =P'_sM_s(P'_{s-1})\cdots M_2(P'_1)M_1&\ket{0^n}
  \end{aligned}
\label{eq:maintext_errorfree_output}
\end{equation}
with $P'_k=P_kP_{k-1}\cdots P_2 P_1$ for $k=1,2,\ldots,s$. Note that the output probability distribution of each pure state $W(P_1,\ldots,P_s)$ can be weakly simulated and the coefficients $c_i(P)$ can be measured experimentally via gate tomography (for single MGs). 
Moreover, errors which occur during intermediate measurements can be similarly taken into account by using the fact that during the computation the qubits are only measured in the computation basis. Hence, only bit--flip errors, which can be applied to the classical output, have to be taken into account (phase--flip errors commute with the measurement). Taking also the measurement errors into account (see Appendix~\ref{appendix:errors-in-measurements}), the output state has a similar form as $\rho_{\rm exp}$ and can therefore be weakly simulated.

Running the verification circuits on the quantum computers gives us a sample $\{y_1,\ldots y_k\}$ that we want to compare with the output $\{z_1,\ldots z_l\}$ of the weak simulation, where $k,l \in poly(n)$. Using then the KS \cite{Ko33} or ES \cite{EppsSingleton} test (see Appendix~\ref{appendix:stat-tests}) allows one to gain confidence that the two samples stem from the same (unknown) distribution by computing a distance between their empirical distribution functions. Up to our knowledge, these tests are among the most widely used tests for the two-sample problem.

{\it The protocols:} Our protocols aim to verify a realization $U'$ of an arbitrary universal quantum computation $U$. To obtain $U'$, one decomposes $U$ into single qubit and $\CZ$-gates and maps those to n.n. MGs and $\CZ$-gates in $U'$. Furthermore, one could implement each $\CZ$ gate in $U'$ by consuming one copy of the magic state $\ket M$ and adaptive measurements. Our protocols can then be summarized as follows:  1. Use one of the options to construct a classically efficiently simulable verification circuit: 1a. Replace each gate $\CZ$ by $\CZ\cdot \SWAP=\FSWAP$ (or any other MG) to obtain a MG circuit. 1b. Consider the realization of the $\CZ$ gates via the magic state $\ket{M}$ and adaptive measurements. Apply the exact same computation to the input state where each magic state $\ket{M}$ is replaced by the state $\ket{M'}=\FSWAP_{2,3}\ket{\Phi^+}_{1,2}\ket{\Phi^+}_{3,4}=\CZ_{2,3} \ket{\Phi^+}_{1,3}\ket{\Phi^+}_{2,4}$ (or any other resourceless state). Note that if there were no errors, both circuits can be simulated classically efficiently and can be directly compared to the output of the quantum computation. 2. Estimate the error parameters $c_i(\cdot)$ in Eq.~(\ref{eq:maintext_rho_out}) for each gate and measurement through process tomography. 3. Run the randomly compiled version of the verification circuits. Due to RC, the output state is given by Eq.~(\ref{eq:maintext_rho_out}). Confidence on the output and error of the whole circuit is gained by comparing the samples obtained from the quantum computer with those obtained via weak classical simulation thereof using e.g. the KS or the ES test.

Before illustrating the performance of these verification protocols, the following comments are in order. First, we assess here the performance of a specific realization $U'$ of the computation $U$ (see Fig.~\ref{fig:results}). Hence, this provides a lower bound on the quality of the quantum device in realizing this particular computation.  Second, one could use other gate replacements to make a circuit (including RC) classically efficiently simulable. For example, one could decompose a circuit into MGs and Hadamard gates, $H$ \cite{Brod2011ExtendingMGs}, and replace the latter with e.g. the MGs $G(H,H)$. Third, to improve the tests and to make the verification circuits even more similar to the original circuit, one can, e.g. use that ${\cal O} (\log(n))$ magic states can still be classically efficiently simulated \cite{Hebenstreit2020}. Moreover, 
the probability with which the final state is in the even parity subspace gives additional information on the error. Fourth, the KS test can be applied to various mappings of the sampled bitstrings to numbers. 
Finally, using the obtained samples, various additional test can be performed. To check whether our protocols could distinguish two error models at all, one can compare the outputs of two entirely classical simulations. This may also be used to determine the necessary sample sizes.

\begin{figure*}[htb!]
    \centering
    \includegraphics[width=\linewidth]{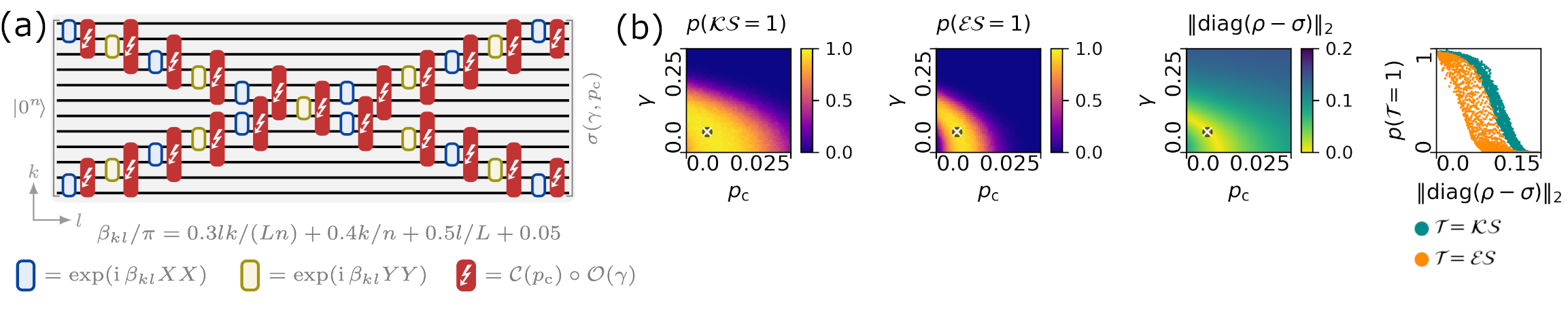}
    \caption{(a) The MG circuit ($n=12$ qubits, depth $L=11$) we use to illustrate our method. To each gate $\exp ( \mathrm{i} \beta H)$, we associate (i) a coherent overrotation $\mathcal{O}$ corresponding to $\exp(\mathrm{i} \gamma \vert \beta \vert H)$ and (ii) a stochastic crosstalk channel $\mathcal{C}$, which is constructed by concatenating in total four channels $\mathcal{\tilde C}(\rho) = (1-p_\mathrm{c}) \rho + p_\mathrm{c}/4 \sum_{P \in \mathcal{X}} P \rho P$, for each pair of targeted qubit and its neighbors, where $\mathcal{X} = \{XX,XY,YX,YY\}$ \cite{HeussenIonErrorModel} (for details and additional examples, see Appendix~\ref{appendix:numerical_results}). 
    In (b), we compare a reference erroneous output state $\rho \equiv \sigma(\gamma = 0.05, p_\text{c} = 0.005)$ (marked by the crosses) with various states $\sigma(\gamma, p_\text{c})$. We find that the relation between the probability of passing the tests, $p(\mathcal{KS}=1)$ and $p(\mathcal{ES}=1)$, and $\Vert \operatorname{diag}(\rho - \sigma)\Vert_2$ behaves similarly across various examples.}
    \label{fig:the-test-circuit}
\end{figure*}

{\it Illustration of the method:} To illustrate the capability of our method to distinguish errors, we compare samples drawn from two classical simulations, one of which takes the place of the quantum computation. We specify the error of the latter and determine how much another set of errors must differ to be distinguished successfully. We consider the MG circuit depicted in Fig.~\ref{fig:the-test-circuit}a, which includes physically relevant crosstalk and overrotation errors. Aiming to analyze the quality of the protocol proposed here, we compute the full density matrices of the output states $\rho$ and $\sigma$, and calculate a distance between their probability distributions $\operatorname{diag}(\rho)$ and $\operatorname{diag}(\sigma)$. With $\mathcal{KS}(\rho, \sigma, M, \alpha)$, we denote the random variable taking the value $0$ if the KS test rejects the null hypothesis $\operatorname{diag}(\rho) \approx \operatorname{diag}(\sigma)$ using $M$ measurement shots with significance level $\alpha$\footnote{The significance level $\alpha$ bounds the probability that the test rejects the null hypothesis, given that it is true.}, and $1$ otherwise. Likewise, we define $\mathcal{ES}$ for the ES test. To estimate $p(\mathcal{KS}=1)$, we repeatedly (here, $1000$ times, using $M=400$ and $\alpha=0.05$) apply the test to newly sampled data and estimate the expectation value $\mathbf{E}(\mathcal{KS}=1)$. In our examples (see Fig.~\ref{fig:the-test-circuit}b and Appendix~\ref{appendix:numerical_results}, where we demonstrate e.g. detection of error drift), the states $\rho$ and $\sigma$ can be distinguished by the statistical tests with high probability as soon as $\Vert \operatorname{diag}(\rho - \sigma) \Vert_2 \geq d_\text{crit} \approx 0.1$. In such cases, the error model is rejected. Note that $d_\text{crit}$ can be improved by e.g. increasing $M$ or applying the circuit twice. To demonstrate the applicability of our methods, we include in Appendix~\ref{appendix:numerical_results} numerical results for different type of circuits (brickwall layout) acting on up to $40$ qubits. We also sometimes find that post-processing the samples increases the power of the tests (see Appendix~\ref{appendix:other-methods}).

In the future, it would be intriguing to see how our protocols can be combined with existing verification approaches, also in the context of fermionic quantum computation~\cite{FermionicErrorMitigation} and quantum simulation. Furthermore, it would be interesting to analyze whether our methods could be useful in the case of multi-qubit measurements in Clifford circuits (see \cite{JS17}), where the output can be weakly simulated  \cite{MaartenJozsa}.  However, symmetries of stabilizer states potentially hiding errors and possibly uniformly random measurement outcomes in the error-free case are obstacles one needs to overcome here.

{\it Code Availability: } Simulation codes and input parameters are available on Zenodo \cite{data_available}.

\begin{acknowledgments}
{\it Acknowledgement:} We thank Richard Jozsa and Richard Kueng for helpful discussions. JC, AN, ML, and BK acknowledge financial support from the Austrian Science Fund (FWF) through the grants SFB BeyondC (Grant No. F7107-N38) and P 32273-N27 (Stand-Alone Project). JC acknowledges financial support from the research project "Munich Quantum Valley", Teilprojekt K8 "Hardware Adapted Theory" and BMBF (MuniQC-Atoms, DAQC). ML and BK acknowledge funding from the BMW endowment fund and the Horizon Europe programmes HORIZON-CL4-2022-QUANTUM-02-SGA via the project 101113690 (PASQuanS2.1) and HORIZON-CL4-2021-DIGITAL-EMERGING-02-10 under grant agreement No. 101080085 (QCFD).
\end{acknowledgments}

\appendix

\section{Matchgate circuits with magic states and adaptive measurements are universal} \label{appendix:mgc-measurements-and-magic-states}
\begin{figure}
  \centering
  \includegraphics[width=\columnwidth]{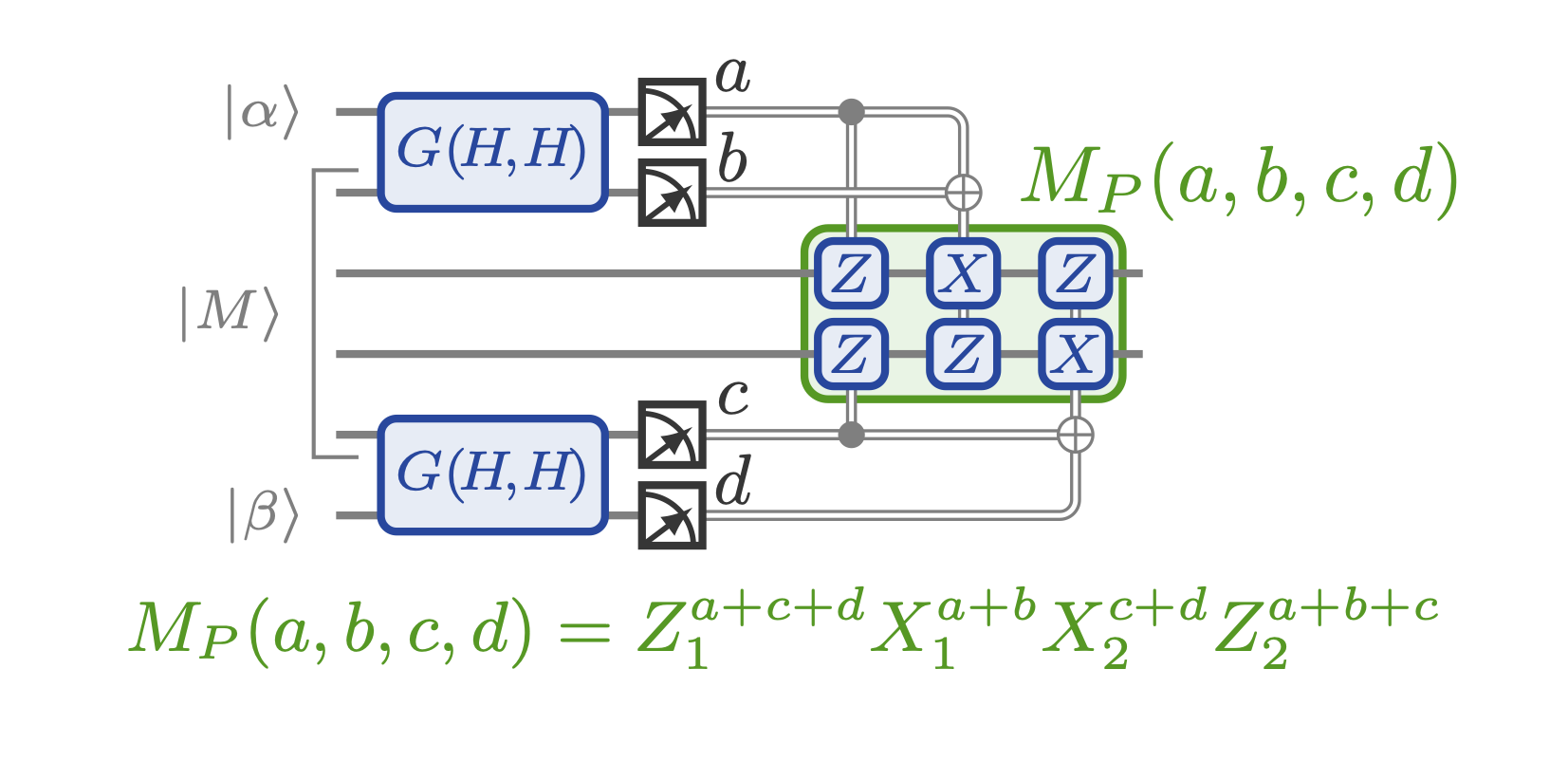}
  \caption{Gadget implementing the resourceful gate
    $\CZ_{k,k+1}$ using a single copy of the magic state
    $\ket{M}=\CZ_{2,3}\ket{\phi^+}_{1,2}\ket{\phi^+}_{3,4}$. As
    usual, the magic state ${\ket M}$ is first placed between the two
    lines of interest using resourceless gates. Then the matchgate
    $G(H,H)$ is applied twice and their output lines are measured to 
    obtain the four bits $a$, $b$, $c$ and $d$. After that, the
    correction Pauli $M_P(a,b,c,d)$ is applied. It is possible
    to implement $M_P(a,b,c,d)$ using only MGs and auxilliary qubits in computational basis states, which need to be moved from a fringe position to the position of the gadget and back~\cite{Hebenstreit2019}.
    Note that the circuit remains classically efficiently simulable if Pauli unitaries are added to the MG group (see App.~\ref{appendix:simulation-of-mgs}). Hence, we can allow for applications of Pauli gates in the gadget.}
  \label{fig.cz}
\end{figure}

Nearest-neighbors matchgates (MG) together with nearest-neighbor $\CZ$ gates form a universal gate set~\cite{JozsaMiyake, Brod2011ExtendingMGs}. We are interested in a realization of $\CZ$ that uses adaptive measurements and consumes one copy of the magic state $\ket M={\sf CZ}_{2,3}\ket{\Phi_+}_{1,2}\ket{\Phi_+}_{3,4}$ (as illustrated in Fig.~\ref{fig.cz}). The purpose of this appendix is to review how to rewrite an arbitrary quantum computation in the latter form. We start with a quantum circuit acting on $\ket{0^m}$ whose gates are either single-qubit unitaries or controlled-$Z$ operations $\CZ_{ij}$, acting on the qubits $(i,j)$. Let $r$ denote the number of gates of the form $\CZ_{ij}$ and $s$ the total number of gates.

First of all, note that the gates $\CZ_{ij}$ can be expressed in terms of at most $2m$ n.n. swap gates $\SWAP_{i,i+1}$ and a single n.n. controlled-$Z$ gates $\CZ_{i,i+1}$. As shown in~\cite{Brod2011ExtendingMGs}, one can define an equivalent quantum circuit acting on $n=2m$ qubits with initial state $\ket{0^{n}}$. The new computation will entirely lie in the so-called \emph{encoded} subspace, the subspace of $\CC^{2^{n}}$ spanned by elements of the form $\ket{x_1x_1x_2x_2\cdots x_mx_m}$ with $x_i=0,1$. In the following, we recall this mapping.

First, it is easy to see that each single-qubit unitary $A_i$ acting at qubit $i$ in the original $m$-qubits circuit, is equivalent to the MG
\[
A_i\mapsto G(A,A)_{2i-1,2i}\,,
\]
acting in the encoded computation. The number of gates of this form will be $s-r$.

Second, each swap gate $\SWAP_{i,i+1}$ swapping qubits $(i,i+1)$ in the original $m$-qubits circuit can be mapped to the following MG sequence:
\[
\begin{aligned}
\SWAP_{i,i+1}\mapsto&\FSWAP_{2i,2i+1}\FSWAP_{2i-1,2i} \times \\
   &\FSWAP_{2i+1,2i+2}\FSWAP_{2i,2i+1}\,.
\end{aligned}
\]
In other words, $\SWAP$ in the encoded subspace, $\ket{x_ix_ix_{i+1}x_{i+1}}\mapsto\ket{x_{i+1}x_{i+1}x_ix_i}$, can be realized with four MG of the form $\FSWAP=G(Z,X)$. Thus, the number of gates of this form will be at most $8rm$.

Finally, each controlled-$Z$ gate $\CZ_{i,i+1}$ acting on qubits $(i,i+1)$ in the original $m$-qubits circuit is mapped to
\[
\CZ_{i,i+1}\mapsto\CZ_{2i,2i+1}.
\]
Consider for simplicity the case $m=2$. Then, the action of controlled-$Z$ in the encoded subspace can be obtained as
\begin{multline*}
\ket{x_1x_1x_2x_2}\mapsto\\(-1)^{x_1x_2}\ket{x_1x_1x_2x_2}=\CZ_{2,3}\ket{x_1x_1x_2x_2}\,.
\end{multline*}

We have thus reviewed how to map the original $m$-qubits computation (composed out of $s=poly(m)$ gates, $r$ of which are controlled-$Z$ gates, $s-r$ being single qubit unitaries) to a new one acting on $2m$ qubits whose gates are either nearest-neighbors MGs (resourceless gates, at most $s+r(8m-1)$ of them) or $CZ_{k,k+1}$ (resourceful gates, $r$ of them). One can realize each gate ${\sf CZ}_{k,k+1}$ by using essentially the gadget of Ref.~\cite{Hebenstreit2019}. As depicted in Fig.~\ref{fig.cz}, the gadget consists of two application of the MG $G(H,H)$, the measurement of four qubits in the standard basis, and finally the application of the adaptive correction $M_P(a,b,c,d)$ (c.f. Fig.~\ref{fig.cz}). It consumes one magic state $\ket{M}=\CZ_{2,3}\ket{\phi^+}_{1,2}\ket{\phi^+}_{3,4}$.

Observe that the magic state $\ket M$ can also be swapped through the circuit just using resourceless gates like $\FSWAP$. Finally, let us stress  that the first pair (and the second pair) of qubits of the magic state $\ket M$ can also be swapped resourcelessly, and independent of the other pair. This implies that $\CZ_{ij}$ with $j\neq i+1$ can be implemented with  a single application of the gadget. That is, the qubits on which the $\CZ$ gate is acting on need not be swapped next to each other before implementing the gate. Note that this might significantly simplify the realization of the procedure in experiments.

Finally, we remark that the gadget can be adapted to allow for a deterministic implementation of the resourceless $\FSWAP$ gate when supplied with the modified state $\ket{M'} = \CZ_{2,3} \ket{\phi^+}_{1,3}\ket{\phi^+}_{2,4}$. For this, the correction operators also need to be modified. Specifically, the new correction operators will be the Paulis $\SWAP\, M_P(a,b,c,d)\, \SWAP$. One can also make use of the modified gadget in our verification protocol. This will have the advantage that the output of the circuit will not be correlated with the intermediate measurement outcomes. Both versions of the gadget can be simulated classically efficiently. 

\section{Randomized compilation for Matchgate circuits} \label{appendix:randomized-compilation}
In this appendix we give details about how standard randomized compilation techniques are adapted to the case of MG circuits. We first consider the error free state, and then proceed to include errors in the gates and measurements.

\subsection{Randomized compilation without errors}
Let us first consider the ideal case of no errors, where randomized compilation amounts to applying the circuit. This allows us to fix the notation we use throughout the rest of this Supplemental Material. We consider a MGC $M_s \ldots M_2 M_1$ on $n$ qubits applied to an input state $\ket{0^n}$. With $\rho_\mathrm{ideal} = \ket \psi \bra \psi$, we denote the output state of this circuit. Ideally, in order to prepare this
state in the lab, one has to implement the unitary evolution
\[
  \rho_{(k-1)}\mapsto\rho_{(k)}=M_k\rho_{(k-1)}M_k^\dagger\,,
\]
for $k=1,\ldots,s$ with $\rho_{(0)}=\ket{0^n}\bra{0^n}$.

As in Ref.~\cite{EmersonRandComp}, we propose the following randomized compilation. At each step
$k=1,\ldots,s$, implement the evolution $\si_{(k-1)}\mapsto\si_{(k)}$
with
\[
  \si_{(k)}=\frac1{|{\mathcal P}_n|}\sum_{P\in{\mathcal
      P}_n}PM_k(P)P\si_{(k-1)}PM_k^\dagger(P)P\,
\]
for $\si_{(0)}=\ket{0^n}\bra{0^n}$. Here, $M(P)$ is defined as the MG
\[
M(P)\equiv PMP\,,
\]
as in the main text. Note that this can be easily done
by first sampling a uniformly random $P\in{\mathcal P}_n$ and then
implementing the sequence of three unitary evolutions associated to $PM_k(P)P$ (i.e., first the unitary evolution $P$, then the unitary evolution $M(P)$, and finally again the unitary evolution $P$). Note that, as in standard randomized compilation, the last layer of Pauli gates appearing in $\si_{(s-1)}\mapsto\si_{(s)}$ before a final measurement need not be implemented physically, as it can be taken into account via classical post-processing.

In the absence of errors, the randomized compilation will produce the
same state $\si_{(s)}=\rho_{(s)}=\rho_{\rm ideal}$. As we show in the next appendix, in the presence of noise, the randomized compilation will produce the state $\rho_{\rm exp}$ of the main text.

\subsection{The case of a noisy MG circuit, without adaptive measurements and noiseless Pauli gates}
Let us now consider erroneous gates and review how randomized compilation drives the errors to stochastic Pauli channels.
Errors in the measurements will be discussed in the next appendix. Here, we assume that one can implement Pauli operators
perfectly. In App.~\ref{sec:noisy_Paulis}, we show how errors in the Pauli
gates can be treated similarly. Hence, errors in the circuit will be
modeled as follows. There are channels $\errch_k$ such that, at each
step $k=1,\ldots,s$, the evolution that is actually implemented in the
lab is $\tau_{(k-1)}\mapsto\tau_{(k)}$ with
\[
  \tau_{(k)}=\frac1{|{\mathcal P}_n|}\sum_{P\in{\mathcal
      P}_n}P\errch_k\big(M_k(P)P\tau_{(k-1)}PM_k^\dagger(P)\big)P
\]
for $\tau_{(0)}=\ket{0^n}\bra{0^n}$. The final state of such an
evolution is the noisy, experimental state
$\rho_{\rm exp}=\tau_{(s)}$.

Due to randomized compiling, one can see how the channels $\errch_k$
are respectively driven into stochastic Pauli channels $\pauch_k$
obtained from $\errch_k$ by twirling under the Pauli group. More
precisely, the Pauli twirl of any channel $\errch_k$ is a stochastic Pauli
channel
\begin{multline*}
  \pauch_k(\cdot)=\frac1{|{\mathcal P}_n|}\sum_{P\in{\mathcal
      P}_n}P\errch_k(P(\cdot)P)P\\
  =\sum_{P\in{\mathcal P}_n}c_k(P)P(\cdot)P
\end{multline*}
for some probability distribution $c_k(\cdot)$ over ${\mathcal P}_n$,
i.e., $c_k(P)\ge0$ for all $P\in{\mathcal P}_n$ and
$\sum_{P\in{\mathcal P}_n}c(P)=1$.

Using $M_k(P)P=PM_k$ in the definition of
$\tau_{(k)}$, we then obtain
\[
  \begin{aligned}
    \tau_{(k)}&=\frac1{|{\mathcal P}_n|}\sum_{P\in{\mathcal
        P}_n}P\errch_k\big(M_k(P)P\tau_{(k-1)}PM_k^\dagger(P)\big)P\\
    &=\frac1{|{\mathcal P}_n|}\sum_{P\in{\mathcal
        P}_n}P\errch_k\big(PM_k\tau_{(k-1)}M_k^\dagger P\big)P\\
    &=\sum_{P\in{\mathcal P}_n}c_k(P)PM_k\tau_{(k-1)}M_k^\dagger P\\
    &=\pauch_k(M_k\tau_{(k-1)}M_k^\dagger)\,.
  \end{aligned}
\]

As this holds for all gates, one can express the final noisy state
$\rho_{\rm exp}=\tau_{(s)}$ as
\begin{equation}\label{eq:rho_noisy}
  \rho_{\rm exp}=\sum_{P_1\in{\mathcal
      P}_n}\!\!\cdots\!\!\sum_{P_s\in{\mathcal
      P}_n}c(P_1,\ldots,P_s)W(P_1,\ldots,P_s)
\end{equation}

where $c(P_1,\ldots,P_s)=c_1(P_1)\cdots c_s(P_s)$ and
\[
  W(P_1,\ldots,P_s)=\ket{\psi(P_1,\ldots,P_s)}\bra{\psi(P_1,\ldots,P_s)}
\]
with
\[
  \begin{aligned}
    \ket{\psi(P_1,\ldots,P_s)}&=P_sM_s\cdots P_2M_2P_1M_1\ket{0^n}\\
    &=P'_sM_s(P'_{s-1})\cdots M_2(P'_1)M_1\ket{0^n}
  \end{aligned}
\]
where we have introduced $P'_k=P_kP_{k-1}\cdots P_2 P_1$.

\subsection{Including errors in the measurements} \label{appendix:errors-in-measurements}
Here, we address the effect of noise in a measurement both on the measurement outcome, as well as the post-measurement state. To model such an error, we consider the following operation: An error channel $\errch$ acting on $\log(n)$ qubits, followed by a "perfect" measurement in the $Z$-basis on qubit $k$ giving an outcome $t\in\{0,1\}$ with corresponding projectors $\Pi_t = (\id+(-1)^t)Z_k)/2$. When performing the measurement on $\rho$, the state after obtaining $t$ thus reads
\[
    \rho^{(t)} = \Pi_t \errch(\rho) \Pi_t.
\]
Again, with a RC procedure, one can again drive $\mathcal{E}$ into a stochastic Pauli channel \cite{EmersonRandComp}. For each realization of the measurement, the procedure is as follows:
\begin{enumerate}
    \item Uniformly sample a Pauli operator $P$ from the $n$-qubit Pauli group and apply it to the state before the measurement.
    \item Perform the measurement and obtain outcome $t$.
    \item Act with $P$ again on the post-measurement state.
    \item If for the $k$-th tensor factor of $P$, $[P]_k$, it holds that $[P]_k \in\{X, Y\}$, flip the measurement outcome $t \mapsto t\oplus 1$, otherwise do nothing.
\end{enumerate}

Let us now verify that this procedure gives a stochastic Pauli channel, followed by an ideal measurement. With $\rho^{(t)}_\mathrm{rc}$, we denote the output state of the procedure after obtaining outcome $t$ (this variable now also includes the cases where originally, $t \oplus 1$ was measured and flipped to $t$). We may write
\begin{align*}
\rho^{(t)}_\mathrm{rc} = \frac{1}{\vert\mathcal{P}_{n}\vert} \!\!\!\sum_{\substack{P\in\mathcal{P}_n\\ [P]_k\in\{\id,Z\}}}\!\!\!\!\!\! & P \Pi_t \errch( P \rho P ) \Pi_t P \quad + \\ 
\frac{1}{\vert\mathcal{P}_{n}\vert} \!\!\!\sum_{\substack{P\in\mathcal{P}_n\\ [P]_k\in\{X,Y\}}}\!\!\!\!\!\! & P \Pi_{t\oplus 1} \errch( P \rho P ) \Pi_{t \oplus 1} P.
\end{align*}
Using $[Z_k, \Pi_t] = 0$ and $X_k \Pi_t X_k = Y_k \Pi_t Y_k = \Pi_{t\oplus 1}$, this can be rewritten to 
\[
\rho^{(t)}_\mathrm{rc} = \frac{1}{\vert\mathcal{P}_{n}\vert} \sum_{\substack{P\in\mathcal{P}_n}} \Pi_t P \errch( P \rho P ) P \Pi_t,
\]
from which one can see that the twirling over the Pauli group drives $\errch$ into a stochastic Pauli channel with coefficients $c(\cdot)$. That is,
\[
\rho^{(t)}_\mathrm{rc} = \sum_{P\in\mathcal{P}_n}c(P) \Pi_t P\rho P \Pi_t.
\]
The probability to obtain outcome $t$ then reads
\begin{align*}
p(t) =& \tr \Pi_t \rho_\mathrm{rc} = \sum_{P \in\mathcal{P}_n} c(Q) \tr \Pi_t P \rho P \\
&= (1-\epsilon) \tr \Pi_t \rho + \epsilon \tr \Pi_{t\oplus 1}\rho,
\end{align*}
where $\epsilon$ is the sum of all $c(P)$ such that the $k$-th tensor factor of $P$ is in $\{X, Y\}$.

Let us add two more comments here. Firstly, this procedure can be extended straightforwardly to the case of multiple (simultaneous) measurements, provided that the corresponding Pauli channels have at most $poly(n)$ many non-vanishing coefficients. Secondly, note that when considering adaptive measurements during a circuit, one has to move away the measured qubits to a fringe position. This can either be done by applying a sequence of $\FSWAP$s, if the measurement outcome is $0$, or a sequence of MGs $G(-Z, X)$, if the measurement outcome is $1$. An alternative strategy is to always flip the measured qubit to $\ket 0$ using the Pauli $X$ (MG simulation techniques still work if $X$ is added to the gate set), and then use only $\FSWAP$s to remove the qubit. The latter strategy avoids any adaptive gate sequence of length $\mathcal{O}(n)$ that is not directly part of the gadget in favor of just a single adaptive gate.

\section{Simulation of the verification circuits acting on \texorpdfstring{$\ket {M'}^{\otimes r}$}{M'}} \label{appendix:simulation-of-mgs}
Recall that one type of verification circuit we propose stems from a circuit in which each resourceful gate is implemented using a gadget, acting however on a MG-generatable input state (hence, we may view the input state as $\ket{0\ldots 0}$ to which an additional layer of error-free MGs is applied). Here, we explain how to weakly simulate such a circuit in the presence of stochastic Pauli noise. Suppose that we have $r$ single-qubit adaptive measurements, which we label by $k=1,\ldots, r$. We denote by $s_k$ the number of MG and measurements (i.e. any operation that causes a stochastic Pauli error after a randomized compilation procedure) before (and including) the $k$th adaptive measurement. That is, the operations $M_{s_{k-1} +1}, \ldots, M_{s_{k}-2}$ are MGs, the $s_{k}-1$-th operation is the $k$-th measurement, and the $s_{k}$-th operation is the correction operator $M_{s_k}(a_k)$ conditioned on the measurement outcome $a_k$.
In Eq.~(\ref{eq:rc_state_with_intermediate_measurements}) at the end of this appendix, we explicitly write down the output state. Before doing so, we first describe an algorithm for a weak simulation of the circuit. To this end, we will use the notation
\[
\begin{aligned}
    \vec P_{s}&=(P_1,P_2,\ldots,P_s)\in(\mathcal P_n)^s\,,\\
    \vec a_{m}&=(a_1,a_2,\ldots,a_{m})\in\{0,1\}^{m}\,.
\end{aligned}
\]
With $W^{\mathrm{(pre)}}(\vec P_{s_k},\vec a_{k-1})$ and $W(\vec P_{s_k}, \vec a_k)$, we denote the normalized states directly before and after the $k$-th intermediate measurement respectively. That is,
\begin{multline*}
    W^{\mathrm{(pre)}}(\vec P_{s_k},\vec a_{k-1})=\\
    U_k(\vec P_{s_k})W(\vec P_{s_{k-1}},\vec a_{k-1})U^\dagger_k(\vec P_{s_k}),
\end{multline*}
where
\begin{multline*}
    U_k(\vec P_{s_k})= P_{s_{k}-2} M_{s_k - 2} 
    \ldots P_{s_{k-1}+1} M_{s_{k-1}+1}\,
\end{multline*}
are the unitary evolution operators between the $k-1$-th and $k$-th measurement. Furthermore,
\begin{multline*}
    W(\vec P_{s_k}, \vec a_k) = \frac{1}{p(\vec P_{s_k},\vec a_{k})} \Big( P_{s_k} M_{s_k}(a_k) \Pi_{a_k} \times \\
    \;\; P_{s_k-1}W^{\mathrm{(pre)}}(\vec P_{s_k},\vec a_{k-1})P_{s_k-1} \Pi_{a_k} M_{s_k}^\dagger(a_k) P_{s_k} \Big),
\end{multline*}
where we introduced the projectors
\[
    \Pi_{a_k}=\frac{\id+(-1)^{a_k}Z_{q(k)}}2,
\]
with $q(k)$ denoting the qubit which is measured at the $k$-th measurement, and the normalization constant $p(\vec P_{s_k},\vec a_{k})$ such that $\tr(W(\vec P_{s_k},\vec a_{k}))=1$\footnote{The case $\tr(W(\vec P_{s_k},\vec a_{k}))=0$ would correspond to a measurement outcome with probability $0$, and as such, we do not need to keep track these states.}. Note that the $s_{k}$-th MG is a {\em correction} MG that depends on the intermediate measurement outcome $a_{k}$. Since for the $\CZ$-gadget, these correction operators $M_{s_k}(a_k)$ are in fact Pauli operators, we do not need to associate error channels with them, as we assume that Paulis can be implemented perfectly. Hence, $P_{s_k}=\id$ (we include it here, as we consider the case of erroneous Paulis in App.~\ref{sec:noisy_Paulis}). The Pauli $P_{s_k-1}$ is associated to errors in the measurement.

To weakly simulate the MG circuit in question, one needs a sampling procedure for a bitstring $x = (x^{(1)},\ldots, x^{(n)})$ at the end of the circuit. Such a procedure can be obtained by making use of the following two subroutines:
\begin{enumerate}
\item Between two measurements $k$ and $k+1$, the Paulis $(P_{s_k},P_{s_k+1},\ldots,P_{s_{k+1}})$ are each sampled according to\footnote{By denoting $X \sim p(X)$, we say that the random variable $X$ takes values $x$ with probabilities $p(x)$.}
\[
    P_i \sim c_i(P_i), i = s_k, s_{k}+1, \ldots, s_{k+1}.
\]
\item For the intermediate measurement $k$, the outcome $a_k$ is sampled according to
\[
a_k\sim\tr(\Pi_{a_k}W^{\mathrm{(pre)}}(\vec P_{s_{k}}, \vec a_{k-1})).
\]
This can be computed efficiently because of the following reason. First, note that 
the states $W^{\mathrm{(pre)}}(\vec P_{s_{k}}, \vec a_{k-1}))$ are constructed by applying MGs, Pauli operators\footnote{Pauli operators can also be treated within the MG formalism. Especially, the classical simulation complexity remains the same if Pauli operators are added to the MG gate set. To see why, recall that for any MG, $M$, and any Pauli, $P$, $PMP$ is again a MG. Thus, any Pauli operator can be pseudo--commuted towards the measurement and then taken into account with post--processing (see also Eq.~(\ref{eq:maintext_errorfree_output})).} and projections in the computational basis. With this, the probability of obtaining $a_k = 0$ or $a_k = 1$ can be calculated efficiently and therefore one can sample the bit $a_k$ \cite{TerhalMG}. 
\end{enumerate}
Producing samples at the end of the circuit can simply be understood as an extension of this scheme: First, one simulates only a measurement of the first qubit and samples an output bit $x^{(1)}$. Next, one considers the post-measurement state conditioned on the outcome and samples a bit $x^{(2)}$ for the second measurement. This procedure is repeated until the last qubit, which then yields the bitstring $x = (x^{(1)},\ldots, x^{(n)})$.

For completeness, we can write down the state $\rho^{(k)}$ after the $k$-th (intermediate) measurement as
\begin{equation}
\begin{aligned}
\rho^{(k)} =\!\!\! \sum_{P_{s_k} \ldots P_{1}} \sum_{a_k, \ldots,a_1} \!\!\! c_{s_k}(P_{s_k}) \ldots c_{1}(P_1) \;\times \\
    p(\vec P_{s_k}, \vec a_k) \ldots p(\vec P_{s_1}, \vec a_1) W(\vec P_{s_k},\vec a_{k}).
    \end{aligned}
    \label{eq:rc_state_with_intermediate_measurements}
\end{equation}

\section{The case of noisy Pauli gates}\label{sec:noisy_Paulis}
In this appendix we analyze the case in which the Pauli gates in the randomized compilation are noisy. As we will see, randomized compilation still works, provided the noise of the Pauli gates is gate-independent \cite{EmersonRandComp}. In this case, one can still sample from the output distribution.

In fact, the output state in the case of noisy Pauli gates is analogous to that of Eq.~\eqref{eq:rho_noisy}, but replacing $c_i(\cdot)\mapsto d_i(\cdot)$, where $d_i(\cdot)$ characterize the concatenation of the noise channels associated to a consecutive MG and a Pauli string. This is essentially the same result as in Ref.~\cite{EmersonRandComp}, but adapted to the case of MG circuits.

To illustrate this, consider two consecutive steps in the randomize compilation procedure, say the first two. We first implement the noisy evolution $P_1M_1(P_1)P_1$, and afterwards $P_2M_2(P_2)P_2$. In the lab, this is done by a sequence of noisy unitary evolutions associated to, the Pauli $P_1$, the MG $M_1(P_1)$, the Pauli $Q_{21}=P_2P_1$, the MG $M_2(P_2)$, and the Pauli $P_2$.

Let $\Delta$ be the noisy channel characterizing the gate-independent noisy implementation of Pauli strings. This is, instead of the unitary map $P(\cdot)P$, we assume one actually implements $P\Delta(\cdot)P$~\footnote{This error model for noisy Pauli gates is the same as the one considered in~\cite{EmersonRandComp}.}. As above, we denote by $\cE_k$ the noisy channel associated to $M_k$.

Consider the noisy implementations of $P_1$, $M_1(P_1)$ and $Q_{21}$. Since the first Pauli $P_1$ can be implemented via post-processing, it can be considered noiseless\footnote{Note that this amounts to assume that any computational basis state can be prepared perfectly.}. The noisy implementation reads
\begin{multline}\label{eq:pnoise}
Q_{21}\Delta\Big(\cE_1\big(M_1(P_1)P_1(\cdot)P_1M_1^\dagger(P_1)\big)\Big)Q_{21}=\\
P_2P_1\Delta\Big(\cE_1\big(M_1(P_1)P_1(\cdot)P_1M_1^\dagger(P_1)\big)\Big)P_1P_2=\\
P_2P_1\Delta\Big(\cE_1\big(P_1M_1(\cdot)M_1^\dagger P_1\big)\Big)P_1P_2.
\end{multline}
After summing over $P_1$, the channel $\Delta\circ\cE_1$ is depolarized the right hand side of Eq.~\eqref{eq:pnoise} reads
\[
P_2\pauch_1\big(M_1(\cdot)M_1^\dagger\big)P_2\,,
\]
where
\begin{multline*}
  \pauch_k(\cdot)=\frac1{|{\mathcal P}_n|}\sum_{P\in{\mathcal
      P}_n}P(\Delta\circ\cE_k)(P(\cdot)P)P\\
  =\sum_{P\in{\mathcal P}_n}d_k(P)P(\cdot)P\,.
\end{multline*}

Note that this reasoning can be iterated. Consider next the noisy implementation of $M_2(P_2)$ and $Q_{32}=P_3P_2$, one obtains 
\[
Q_{32}\Delta\bigg(\cE_2\big(M_2(P_2)P_2\pauch_1\big(M_1(\cdot)M_1^\dagger\big)P_2M_2^\dagger(P_2)\big)\bigg)Q_{32}\,,
\]
and after summing over $P_2$ one has
\[
P_3\pauch_2\bigg(M_2\pauch_1\big(M_1(\cdot)M_1^\dagger\big)M_2^\dagger\bigg)P_3\,.
\]
It is easy to see that the final outcome of such a procedure is given by Eq.~\eqref{eq:rho_noisy}, although replacing $c_i(\cdot)\mapsto d_i(\cdot)$.
Let us finally mention that the case where the error depends on the Pauli gate has been studied in \cite{EmersonRandComp} (see Theorem 2 in \cite{EmersonRandComp} for the case where the easy gates are Pauli gates).

\section{Comparing samples: two statistical tests} \label{appendix:stat-tests}

Here, we review two well-known two-sample tests from classical hypothesis testing. With such tests, the aim is to gain insight on whether two samples $x_1, \ldots, x_k$ and $y_1, \ldots y_l$, originate from the same probability distribution or not. That is, with $x_i \sim p^{(1)}(x_i)$ and $y_i \sim p^{(2)}(y_i)$, the question becomes: Is $p^{(1)} = p^{(2)}$ or $p^{(1)} \neq p^{(2)}$? A strategy that several tests make use of is to assume that the so-called \emph{null hypothesis} $p^{(1)} = p^{(2)}$ is true, and then calculate some test statistic $T(x_1, \ldots, x_k, y_1, \ldots, y_l)$ with a known distribution. Ideally, this distribution is independent of $p^{(1)}$. This allows one to specify a \emph{significance level} $\alpha \in (0,1)$, and a set of critical values $\mathcal{T}_\alpha$, such that if $T \notin \mathcal{T}_\alpha$, one would reject the null hypothesis. The crucial part is to choose the set $\mathcal{T}_\alpha$ in such a way that the probability of rejecting the null hypothesis, given that it is in fact true, is bounded with
\begin{equation}
    \prob(T \notin \mathcal{T}_\alpha | \, p^{(1)} = p^{(2)}) \leq \alpha.
    \label{eq:bound_on_error_test}
\end{equation}
The quantity $1-\alpha$ is known as the \emph{confidence level}. On the other hand, for the tests we use, the probability $\delta$ of keeping the null hypothesis, even though it is wrong, cannot be estimated generically (that is, provided $p^{(1)}$ and $p^{(2)}$ are unknown). The quantity $1 - \delta$ is known as the \emph{power} of the test.

In the following, we review two established two-sample tests, namely the Kolmogorov-Smirnov (KS) and the Epps-Singleton (ES) test. Whereas the KS test has mostly been studied for continuous distributions, the case where the distributions are discrete has been analyzed in \cite{WalshKSDiscrete}. We apply in our examples the KS test implemented in the software package \texttt{scipy}, version \texttt{1.8.0} \cite{ScipyRef} and observe that the bound in Eq.~\eqref{eq:bound_on_error_test} works reasonably well. Additionally to that, we use the ES test, which has been developed specifically to account both for continuous and discrete distributions.

\subsection{Kolmogorov-Smirnov test}
Kolmogorov and Smirnov provided one of the original ideas to tackle the problem of comparing samples. Their methods work both for the one-sample (which is to decide whether a sample has a given distribution) and two-sample problem. Here, we first discuss the former and later comment also on the two-sample problem. Let $F(x)$ be the ideal cumulative distribution function\footnote{The CDF $F$ of a (discrete) random variable $X$, taking values $x$ with probabilities $p(x)$, is defined as $F(x) = \sum_{t\leq x}p(t)$. The \emph{empirical} CDF of a sample $x_1,\ldots,x_k$ is thus $F_k(x) = \frac{1}{k} \# \{l : x_l \leq x\}$.} (CDF) and $F_{k}$ be the empirical CDF for $k$ observations. Under the null hypothesis that the observations are distributed according to $F(x)$, the Kolmogorov-Smirnov test makes use of the behaviour of the statistic
\[
d_k \equiv {\rm sup}_x|F_k(x)-F(x)|
\]
as the number of observations $k$ grows. When $F(x)$ is continuous, the limiting distribution of the random variable $d_k$ is distribution-free (independent of $F(x)$ under the null hypothesis) and called the Kolmogorov distribution. Intuitively, $d_k$ goes to zero as the number of observations $k$ grows to infinity. When $F(x)$ is continuous, Kolmogorov computed rigorously the asymptotic distribution of $d_k$~\cite{Ko33} and later on Smirnov published a table with the corresponding critical values~\cite{Sm48}. This is a very typical goodness of fit test: (i) set a significance level $\alpha$, (ii) using the published tables, find the critical value $d(\alpha)$ such that $\prob(d<d(\alpha)) \geq 1-\alpha$, (iii) estimate $d_k$ from $k$ observations and reject the null hypothesis if $d_k>d(\alpha)$. Importantly, the computation of $d_k$ is efficient in $k$ and tables of critical values $d(\alpha)=d_k(\alpha)$ are available. However, this is not the situation we are dealing with. In fact, we want to test whether two samples $\{s_1,\ldots,s_k\}$ and
$\{r_1,\ldots,r_l\}$ come from the same probability distribution or not. In other words, we do not have access to the ideal CDF $F(x)$, we have access to two empirical CDFs $F_{k,1}(x)$ and $F_{l,2}(x)$. For instance, $F_{k,1}(x)$ could correspond to the empirical CDF computed from $k$ measurements of the noisy quantum device, and $F_{l,2}(x)$ is the empirical CDF computed from $l$ repetitions obtained with our simulation scheme. One can still use a variant of the Kolmogorov-Smirnov test in this case by modifying the statistic to
\[
d_{k,l} \equiv {\rm sup}_x|F_{k,1}(x)-F_{l,2}(x)|.
\]
Tables with the corresponding critical values $d_{k,l}(\alpha)$ are available (for instance in \cite{knuth97}).

Since we aim to apply the test to samples of bitstrings, we remark here again that there is an assumption on the continuity of $F_{k,1}$ and $F_{l,2}$ (which is made use of when showing that the test statistic is distribution-free). As mentioned before, in \cite{WalshKSDiscrete} the test has been extended to discrete distributions for both the one- and two-sample problem. For an alternative treatment of the problem, we also use the ES test.

\subsection{Epps-Singleton test}\label{sec:ES_test}

Several statistical tests, such as the KS test, derive their test statistic from the CDF. Here, we review the Epps-Singleton (ES) test as introduced in \cite{EppsSingleton}, which instead uses a test statistic based on the characteristic function of the involved distributions. One advantage of the ES test is that it can be rigorously applied to discrete samples.

Recall that for a given (continuous or discrete) probability distribution, the characteristic function is defined as the Fourier transform of the density $p$, namely 
\[\phi (t) = \int \mathrm{e}^{\mathrm{i} t x} p(x) \mathrm{d}x. \] 
The objective of the ES test is again: Given two independent samples $x^{(1)}_1, \ldots, x^{(1)}_{N_1} \sim p^{(1)}$ and $x^{(2)}_1, \ldots, y^{(2)}_{N_2} \sim p^{(2)}$, decide whether the null hypothesis $p^{(1)} = p^{(2)}$ or the alternative $p^{(1)} \neq p^{(2)}$ is true. For the ES test, this is reformulated in terms of the respective characteristic functions, i.e. the null hypothesis is $\phi^{(1)} = \phi^{(2)}$.

To construct the test statistic, the initial step is to evaluate the real and imaginary part of the two empirical characteristic functions at $J$ points $t_j, j=1,\ldots, J$ (we comment on a specific choice below). Using the notation 
\begin{align*} g(x) &=\\
(&\cos{(t_1 x )},\sin{(t_1 x )}, \ldots, \cos{(t_J x )},\sin{(t_J x )})^{\mathrm{T}}, \end{align*}
one obtains two vectors 
\[g_{(i)} = \frac{1}{N_i} \sum_{k=1}^{N_i} g(x^{(i)}_k) \in \mathbb{R}^{2J}.\]
Consider now the quantity $G = g_{(1)} - g_{(2)}$ and assume the null hypothesis to be true. From the central limit theorem it follows that for large $N_1, N_2$, the quantity $\sqrt{N_1 + N_2} G$ is asymptotically normally distributed, with mean $\vec 0$ and covariance matrix $\Omega$. Let $r$ and $\Omega^+$ denote the rank and the pseudoinverse of $\Omega$. Defining a statistic $W = (N_1 + N_2) G^\mathrm{T} \Omega^+ G$, one can show that $W$ has asymptotically a chi-squared distribution\footnote{The chi-square distribution with $r$ degrees of freedom is defined to be the distribution of a random variable $Q=\sum_{i=1}^{r} N^2_i$, where the $N_i$ are independent normally distributed random variables with a mean of $0$ and a variance of $1$.} with $r$ degrees of freedom. Notably, this is independent of the distributions of the samples. Thus, it is a good quantity to use as the test statistic. As the covariance matrix is however not known, it remains to construct an estimator for $\Omega$. Epps and Singleton use the estimator
\[\hat{\Omega} = \frac{N_1 + N_2}{2}\left(\frac{1}{N_1} + \frac{1}{N_2} \right)(\hat S_1 + \hat S_2), \]
where
\[\hat S_i = \frac{1}{N_i} \sum_{i=k}^{N_i} g(x^{(i)}_k) g(x^{(i)}_k)^\mathrm{T} - g_{(i)} g_{(i)}^\mathrm{T}.\]

As for the values of $J$ and the $t_j$, Epps and Singleton suggest $J=2, t_1 = 0.4/\tilde \sigma$ and $t_2=0.8 /\tilde \sigma$, where $\tilde \sigma$ denotes the semi-interquartile range\footnote{The semi-interquartile range of a sample $x_1,\ldots,x_k$, which is ordered such that $x_1 \leq x_2 \leq \ldots \leq x_k$, is defined as $\frac{1}{2}(x_{3k/4} - x_{k/4})$.} of the combined sample. These values have been calibrated such that the test performs optimally in several simulations\footnote{Specifically, Epps and Singleton produce a total of 30 pairs of samples from nine families of distributions (comprising normal, uniform, Cauchy, Laplace, symmetric stable, gamma, Poisson, binomial and negative binomial distributions) with various parameters and optimize $J$ and $t_j$ to give a good average power against the alternatives constructed this way.} conducted by the authors \cite{EppsSingleton}. Note that one could in principle optimize the values of $t_j$ to increase the power of the test in distinguishing members in a given family of error models. In their simulations, Epps and Singleton have also compared their test to several other well-known tests, specifically, the KS, the Anderson-Darling (AD) and the Cram{\'e}r-von-Mises (CM) test. They conclude that in most cases, the KS test is outperformed by the other tests. On the other hand, there are cases in which the AD and CM test have higher power than the ES test and vice-versa.

In our numerical investigations (App.~\ref{appendix:numerical_results}), we typically identify the ES test as more powerful. There are however cases in which the KS test performs better. Thus, we suggest comparing both tests using two classical simulations, and then using the test that gives higher power against a plausible set of alternatives for comparing the classical and quantum computation.

\section{Other methods} \label{appendix:other-methods}

Note that in principle, applying the statistical tests to the bitstrings should test the whole output distribution for equality. However, the sensitivity of the test depends on how the bitstrings are mapped to integers. For instance, the mapping, $x = (x^{(1)},\ldots, x^{(n)}) \mapsto \sum_{i=1}^n 2^i x^{(i)}$, which we use in our examples, might not lead to the most powerful test. Hence, it might be advantageous to run the test for various different mappings of bitstrings to integers (e.g. by an additional reordering of the bits). Moreover, on can consider various other tests to compare the samples, e.g. by estimating moments or applying statistical tests to the distribution of other observables of the output state rather than the bitstrings themselves. This might in particular be relevant, in case the computational task is to estimate the expectation value of an observable as is the case in quantum algorithms such as VQE. Below, we give an example of an observable which turns out to be very powerful for the examples studied in App.~\ref{appendix:numerical_results}. Let us also mention here, that in case the noise is sufficiently small, it might be very hard to differentiate the output from a noise-free version of the circuit. To amplify the noise, one could then consider concatenating the circuit several times.

\subsection{An additional test based on post-processing}\label{appendix:postprocessing-for-higher-test-power}

\begin{figure*}[ht!]
    \centering
    \includegraphics[width=\linewidth]{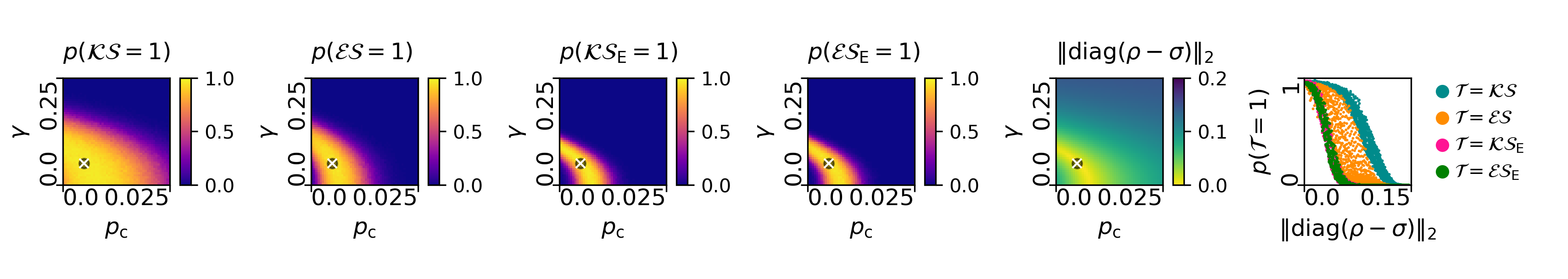}
    \caption{We present the same data as in Fig.~\ref{fig:the-test-circuit}b, however, also including the case where the data has been post-processed with the method described in App.~\ref{appendix:postprocessing-for-higher-test-power} (labeled by the subscript $\mathrm{E}$). In most of our simulations, this post-processing gives an advantage in distinguishing the output states.}
    \label{fig:appendix-energy-method}
\end{figure*}

We elaborate here on how the post-processing of the samples can lead to additional tests and demonstrate that they can give higher power in distinguishing error models in our examples. The idea is that the post-processing makes the statistical test more sensitive to certain differences in the distributions of the output states. In our standard approach, we map bitstrings $x=(x^{(1)},\ldots, x^{(n)})$ to integers using functions of the form $x \mapsto \sum_{i=1}^n x^{(i)} 2^i$. Here, we propose to use the mapping $f(x) = \braket{x}{H_\mathrm{out} \vert x}$, where $H_\mathrm{out} = U H_\mathrm{in} U^\dagger$, $H_\mathrm{in} = -\sum_{i=1}^n Z_i$ and $U$ is the error free MG circuit in question (we recall below why this mapping can be calculated efficiently). The statistical test will then be applied to the samples $\{f(x_1),\ldots, f(x_k)\}$ and $\{f(y_1), \ldots, f(y_l)\}$. In Fig.~\ref{fig:appendix-energy-method}, we present the same example as in the main text, but also apply the tests to the post-processed data. Let us mention that with this construction, if the input state to the circuit is $\ket{0^n}$, then the error free output state $U\ket{0^n}$ will be the unique ground state of $H_\mathrm{out}$. Hence, if $\rho$ is the noisy output state, the expectation value $\tr (H_\mathrm{out}\rho)$, which can be efficiently estimated using $poly(n)$ Pauli measurements, could also be used as an indication of how noisy the circuit is. In principle, one can use any other mapping $f$ and numerically test whether the post-processed samples can be distinguished better.

We now recall why the mapping $x \mapsto \braket{x}{H_\mathrm{out} \vert x}$ can be calculated efficiently, given a description of $U$ in terms of n.n. MGs (see, for instance, \cite{JozsaMiyake}). For this, note that one can write $H_\mathrm{out}$ in terms of the $2n$ Jordan-Wigner operators\footnote{One may define the $2n$ Jordan-Wigner operators as $c_{2k-1} = Z^{\otimes (k-1)}\otimes X \otimes \id^{\otimes(n-k)}$ and $c_{2k} = Z^{\otimes (k-1)}\otimes Y \otimes \id^{\otimes(n-k)}$. A MG circuit $U$ conjugates them via $U c_i U^\dagger = \sum_{j=1}^{2n} R_{ij} c_j$, where $R$ is an orthogonal matrix.} $\{c_i\}$, namely $H_\mathrm{out} = \mathrm{i} \sum_{i,j=1}^{2n} h_{ij} c_i c_j$, where the coefficients $h_{ij}$ can be calculated efficiently. For any computational basis state $\ket x$, $\braket{x}{H_\mathrm{out} \vert x}$ will be a sum over $n$ contributions of the form $\bra x c_{2i-1}c_{2i} \ket x \propto \bra x Z_i \ket x$ (all the other contributions vanish), which thus also can be calculated efficiently.

\section{Further numerical investigations} \label{appendix:numerical_results}

In this appendix, we present the results of some additional selected examples. The aim is to illustrate the capability of our method in several scenarios. This includes the case where the circuit in question corresponds to the encoded version of a circuit on $m=n/2$ qubits, and a demonstration that our method is capable of detecting drifts in the errors. Finally, we also investigate the usefulness of the statistical test for distinguishing different types of random states. Note that one could use such simulations to estimate, or even optimize, the power of the statistical tests against plausible alternatives to the error model. For the statistical tests, we rely on an implementation in the software package \texttt{scipy}, version \texttt{1.8.0} \cite{ScipyRef}.

Let us emphasise that such numerical simulations, where one compares the output of two entirely classical simulations with different error models, are also very useful while verifying quantum circuits. Specifically, it allows to map out in advance many properties of the statistical tests. To give some examples, one can determine the probability that the tests reject the hypothesis of two output states being the same given that the two error models are truly distinct (recall that, generically, one cannot bound the power of the KS or ES test). This allows one to determine which error models can be distinguished at all by the statistical tests and which test performs best for a specific quantum computation. Furthermore, one can optimize the number of measurement shots $M$ as well as test parameters (e.g. the values $t_1,\ldots,t_J$ for the ES test, see App.~\ref{sec:ES_test}).

Let us also comment here on two cases in which the output states are too similar, and hence also cannot be distinguished easily by our methods (specifically, the error parameters on a given order of magnitude may differ vastly, whilst the two output states would be roughly the same).
Firstly, when the errors are too large, the output states become completely depolarized. What matters here is that if two distinct error models give rise to completely depolarized states, both would be valid descriptions for any practical purpose. Secondly, if the errors are very small, the output states are too similar for the size of circuits that we consider in our example. The second case is avoidable by considering sufficiently large circuits, which can be artificially achieved by for instance concatenating a given circuit several times. The minimum distance $\Vert \operatorname{diag}(\rho - \sigma) \Vert_2$, that two output states $\rho$ and $\sigma$ need to have to be distinguished with high probability (which is roughly $10^{-1}$ for the examples we present here) can furthermore be improved by increasing the number of measurement shots $M$.

\subsection{Details on the error model in our examples} \label{appendix:details-on-error-model}

When constructing our examples, we choose an error model that is both gate-dependent, bears physical relevance, but also has an easy parametrization. With this, we consider two types of errors that occur after each gate, namely a crosstalk model that acts on the targeted and neighboring qubits of each gate, as well as an overrotation that acts only on the qubits targeted by a gate.

First of all, let us describe in a bit more detail the stochastic crosstalk channels we use. We choose a model that has been used to describe such errors on an ion trap quantum computer \cite{HeussenIonErrorModel}. Here, the authors have benchmarked a device and found that crosstalk occurs mostly between neighboring qubits. A good average description between two crosstalking qubits $t$ and $b$ is given by the channel
\begin{multline*}
    \mathcal{C}_{tb}(\rho) = (1-p_\mathrm{c}) \rho + \frac{p_\mathrm{c}}{4} \big( X_t X_b \rho X_t X_b \\
    + X_t Y_b \rho X_t Y_b  + Y_t X_b \rho Y_t X_b + Y_t Y_b \rho Y_t Y_b  \big).
\end{multline*}
For instance, a gate applied to qubits $2$ and $3$ would thus cause a crosstalk error of $\mathcal{C}_{21} \circ \mathcal{C}_{23} \circ \mathcal{C}_{32} \circ \mathcal{C}_{34}$. We choose the value $p_\mathrm{c}= 5 \times 10^{-3}$, which is larger than a value of roughly $10^{-4}$ given in \cite{HeussenIonErrorModel}.
The reason for doing so is that the smaller value does not cause a significant deviation in the output states of the small circuits we consider.
We however note that when considering circuits with larger depth, also small errors will cause a deviation in the output state that we could detect with our protocol.

Let us now describe the overrotation errors. These errors are chosen to be gate-dependent, as we would also like to assess the performance of the protocol in such a case. For a gate $\exp(\mathrm{i} H)$, where the generating Hamiltonian is a weighted sum of the Paulis in $\mathcal{Y} = \{XX,XY,YX,YY,Z_1,Z_2\}$, i.e. $H=\sum_{P\in\mathcal{Y}} \beta_P P$,  we use a coherent overrotation with generating Hamiltonian $\gamma \sum_{P\in\mathcal{Y}} \vert \beta_P \vert P$. We choose the value of $\gamma$ such that the average gate fidelity of the error channel with typical values of $\beta_P$ is in the order of magnitude of $0.025$, which has  been recognized as a typically appearing error of two qubit gates on the ion trap quantum computer \cite{HeussenIonErrorModel}. With this, our value of $\gamma$ is $\gamma = 0.05$ (see Fig.~\ref{fig:the-test-circuit}a).

\subsection{Verification of an encoded circuit}

\begin{figure*}[htb!]
    \centering
    \includegraphics[width=\linewidth]{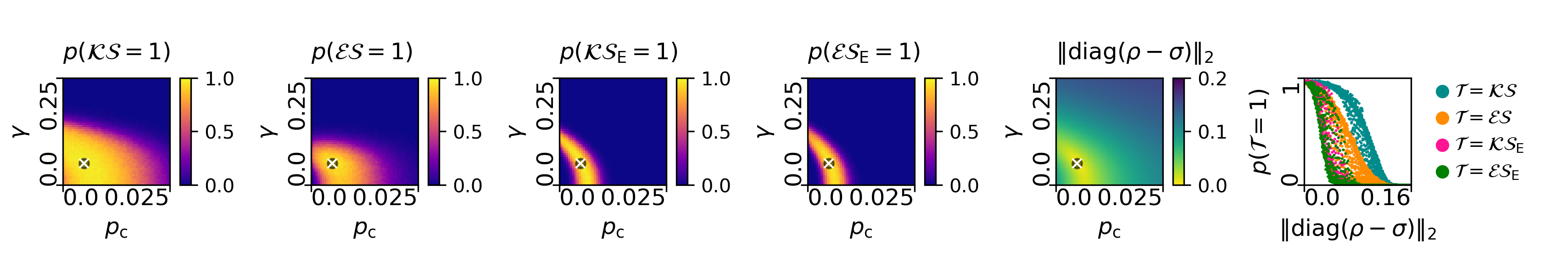}
    \caption{Here, we present the data with the general setup being similar as in Fig.~\ref{fig:the-test-circuit}b. In the circuit however, each $\exp{(\mathrm{i} \beta YY)}$ gate is replaced with an $\FSWAP$ gate. That way, we obtain a circuit that could appear after the encoding and gate replacement steps of our protocol. Again, the here shown results are representative for the instances considered in this work: With an increasing value of $\Vert \operatorname{diag}(\rho - \sigma) \Vert_2$, the probability of correctly distinguishing different errors approaches unity.}
    \label{fig:appendix-fswap-circuit}
\end{figure*}

In order to consider an instance of a verification circuit derived from a universal quantum computation, we proceed as follows. We start with the circuit in Fig.~\ref{fig:the-test-circuit}a, where however each of the $\exp (\mathrm{i} \beta YY)$ gates is replaced with an $\FSWAP$ gate. Recall that in our protocol, $\FSWAP$ gates are the replacements of resourceful $\CZ$ gates. Thus, this new circuit can be considered as an example that would stem from our protocol. The original computation (before encoding) on $m = n/2 = 6$ qubits would then consist of $12$ single qubit $\exp(\mathrm{i} \beta X)$ and $9$ CZ gates. The layout of this circuit can be obtained by grouping together pairs of two qubits in the circuit depicted in Fig.~ref{fig:the-test-circuit}a. We present our data, using various statistical tests, in Fig.~\ref{fig:appendix-fswap-circuit}.

\subsection{Drift in the error parameters} \label{appendix:drift_in_error_parameters}

\begin{figure}[htb!]
    \centering
    \includegraphics[width=\linewidth]{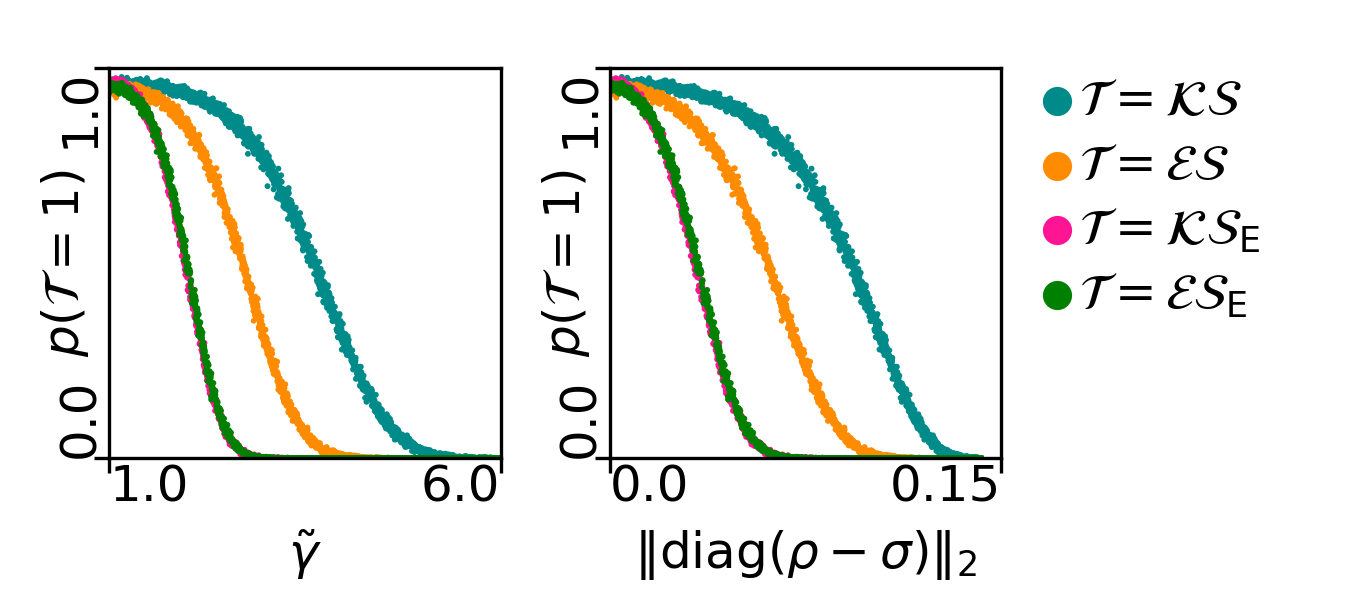}
    \caption{Comparing the outputs of a circuit with static errors to the one of a circuit with layer-dependent overrotation, as given in Eq.~(\ref{eq:app_lin_ovr}). Similar to the results in Fig.~\ref{fig:appendix-energy-method}, a sufficiently large deviation in the output state can be detected by the statistical tests. With this, one is then capable of asserting that a drift-free error model, obtained by process tomography, does not describe the errors that occur in the actual computation.}
    \label{fig:drift_yy}
\end{figure}

In a realistic quantum device, errors might not stay constant with time. An important task would then be to detect whether such drift takes place or not. In other words, one needs to be able to distinguishing the output of a circuit in which no change of error parameters occurs to the output of a circuit with changing errors. Here, we demonstrate that the protocol we propose could be used for such a purpose, provided the distance of the output states becomes large enough. We again compare data from two classical simulations, again with the circuit given in Fig.~\ref{fig:the-test-circuit}a.
The error of one circuit is still characterized by the two parameters $\gamma$ and $p_\mathrm{c}$. For the other circuit, we investigate two different scenarios. In the first one (see Fig.~\ref{fig:drift_yy}), we consider an overrotation that increases linearly with the layer $l$, i.e., the overrotation parameter $\gamma(l)$ is of the form
\begin{equation} \label{eq:app_lin_ovr}
    \gamma(l) = \gamma \cdot \big( 1 - \frac{l}{L} +  \frac{l}{L} \tilde \gamma\big), l=0,\ldots, L-1
\end{equation}
where $L$ denotes the number of layers in the circuit.

\begin{figure}[htb!]
    \centering
    \includegraphics[width=\linewidth]{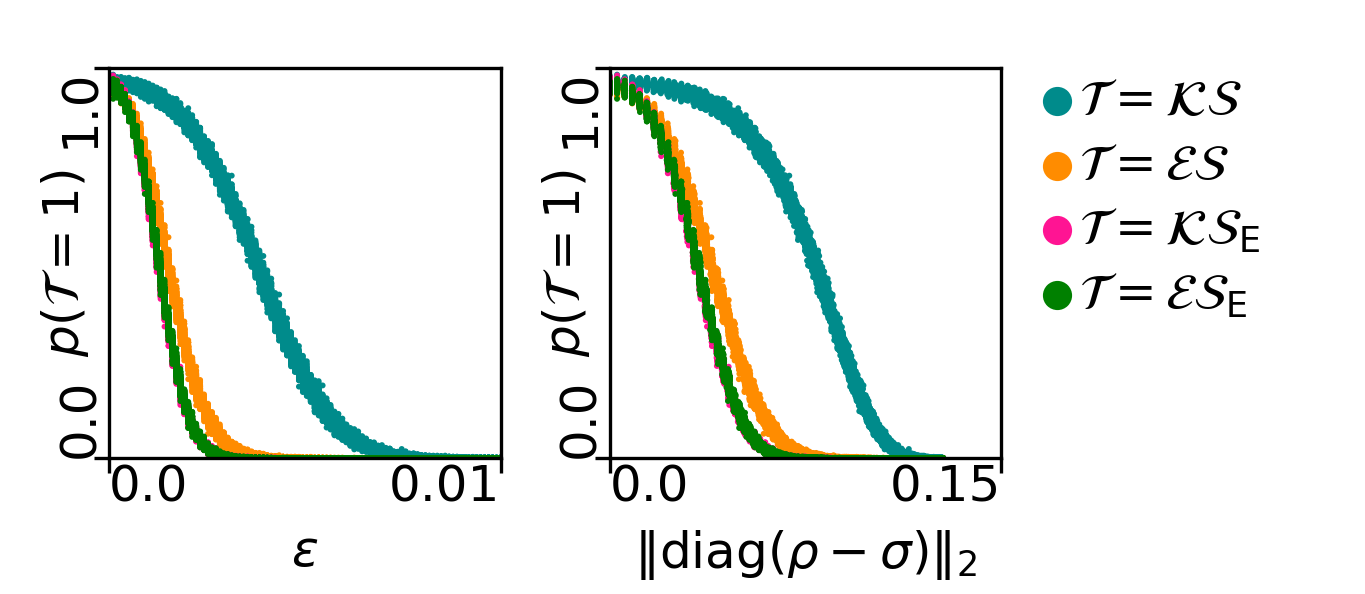}
    \caption{Comparing the output of two circuits with Pauli channel parameters $\{c_k(\cdot)\}$ and $\{\tilde c_k(\cdot)\}$ respectively. For each comparison, the coefficients $\tilde c_k(\cdot)$ are obtained via sampling in an $\epsilon$-ball around reference coefficients $\{c_k(\cdot)\}$ for each gate (see Eq.~(\ref{eq:app_random_pauli})). Similar to Figs.~\ref{fig:appendix-energy-method} and~\ref{fig:drift_yy}, an increasing distance $\Vert \operatorname{diag}(\rho - \sigma) \Vert_2$ between the output states $\rho$ and $\sigma$ increases the probability of detecting deviations in the errors.}
    \label{fig:random_pauli_sampling_yy}
\end{figure}

For the second scenario (Fig.~\ref{fig:random_pauli_sampling_yy}), we ask how different the errors of the gates applied in the actual circuit need to be from the errors obtained via process tomography to be detectable by our protocol. To model this scenario, we consider two sets of Pauli coefficients $\{c_k(\cdot)\}$ and $\{\tilde c_k(\cdot)\}$ respectively. The set $\{c_k(\cdot)\}$ is given, in the same fashion as above, from computing the stochastic Pauli channels stemming from randomly compiled overrotation and crosstalk channels. On the other hand, the set $\{\tilde c_k(\cdot)\}$ is sampled in the vicinity of $\{c_k(\cdot)\}$ as follows: First, we choose an $\epsilon > 0$. Then, for each gate $k$ in the circuit, we construct the perturbed coefficients via
\begin{equation}
\tilde c_k(P) = 
    \begin{cases} \frac{1}{N} \vert c_k(P) + \epsilon \mathcal{N}_P \vert ,
        & \exists Q \text{ s.t. } c_k(Q) \neq 0 \;\wedge \\ & \operatorname{supp}(P) \subseteq \operatorname{supp}(Q), \\
        0, & \mathrm{otherwise},
        \end{cases}
        \label{eq:app_random_pauli}
\end{equation}
where $\mathcal{N}$ is a uniformly random unit vector and $N$ a normalization constant. That is, the new coefficients describe an error channel that acts non-trivially on exactly the qubits on which the original channel has acted on. 

For both scenarios, we find that the statistical tests can distinguish the output states, provided the distance in the respective probability distribution they induce becomes large enough. The qualitative behavior of e.g. $p(\mathcal{KS} = 1)$ w.r.t. $\Vert \operatorname{diag}(\rho - \sigma) \Vert_2$ is very similar in both cases (Figs.~\ref{fig:drift_yy} and~\ref{fig:random_pauli_sampling_yy}), and also to the case where error models with only two error parameters are compared (see Fig.~\ref{fig:the-test-circuit}b text or Fig.~\ref{fig:appendix-energy-method}).

\subsection{Verifying brickwall circuits}

\begin{figure*}[ht]
    \centering
    \includegraphics[width=0.8\linewidth]{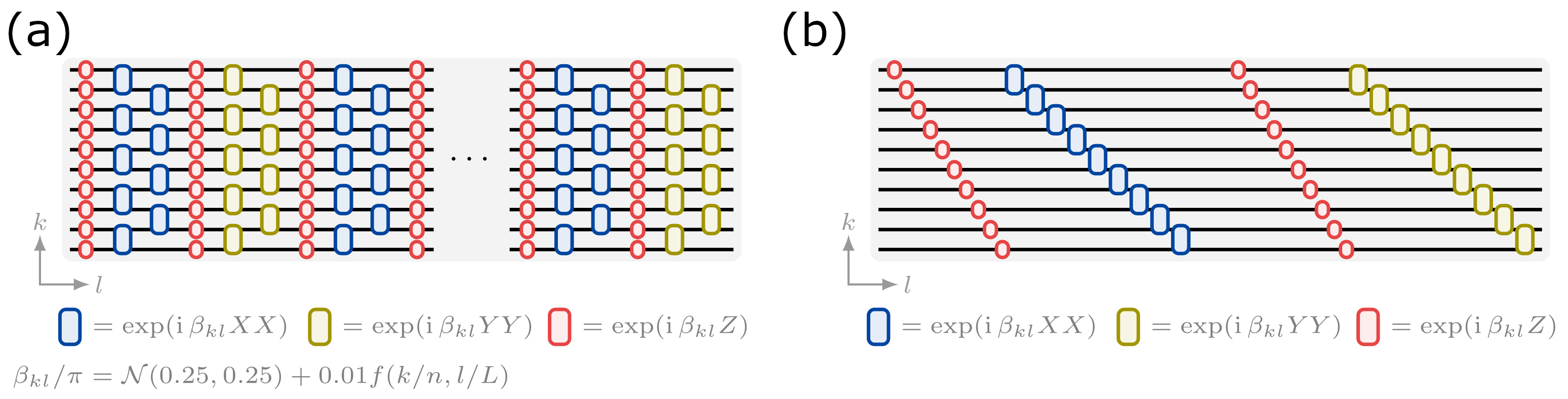}
    \caption{(a) A brickwall circuit layout (here, on $n=10$ qubits with $L=30$ layers) we use to conduct additional testing of our protocols. The gate parameters $\beta_{kl}$ are given by $\beta_{kl} = \mathcal{N}(0.25,0.25) + 0.01 f(k,l)$, where $\mathcal{N}(\mu, \sigma)$ denotes a random variable with mean $\mu$ and variance $\sigma^2$ that is realized independently for each gate. Furthermore, the function $f$, with $0\leq f\leq 1$, is given by $f(k,l) =kl/(n(L-n)) + k/n + l/(L-n)$, where $k=0,\ldots,n-1$ is the index of the first qubit line the gate acts on, and $l=0,\ldots,L-n-1$ is a variable that is increased by $1$ whenever the type of gate changes. For instance, $l=0$ for the first layer of $\exp(\mathrm{i}\beta_{kl}Z)$ gates, then $l=1$ for the second two layers of $\exp(\mathrm{i}\beta_{kl}XX)$ gates and so on. In order to treat errors separately for each gate, we simulate instead a version of the circuit in which the gates are implemented sequentially. That is, the resulting circuit is composed of several concatenations of the circuit depicted in (b). Crucially, in the case of no errors, both circuits coincide in their action.}
    \label{fig:brickwall_layout}
\end{figure*}

Finally, we investigate the capabilities of our protocol when applied to circuits comprising a much larger number of gates. To do so, we extend our numerical simulations to a family of structured circuits, typically referred to as brickwall circuits (see Fig.~\ref{fig:brickwall_layout}). Circuits of such a layout are often the subject of study in various verification and benchmarking protocols (see e.g. \cite{FeKa19, MirrorCircuitsTrust,MirrorRB,KnillRB}). The gate-dependent errors are the same as in the previous examples.

\begin{figure*}[htb!]
    \centering
    \includegraphics[width=0.9\linewidth]{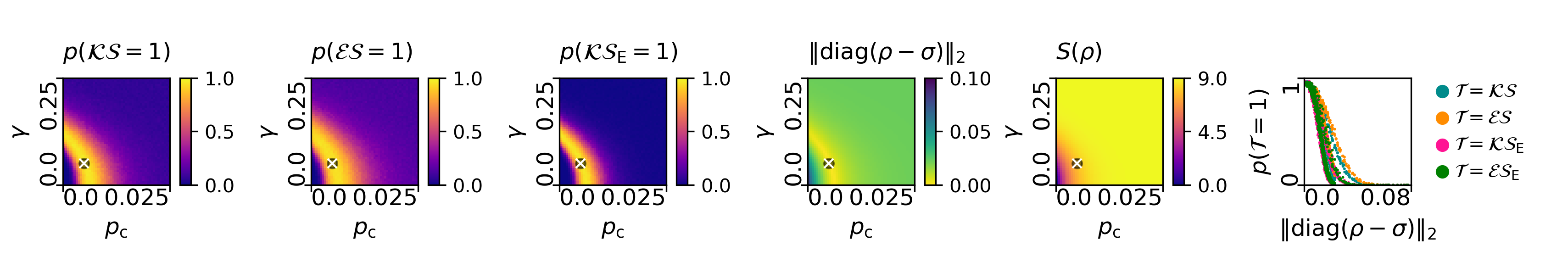}
    \caption{Comparing the output states of a random brickwall circuit (see Fig.~\ref{fig:brickwall_layout}) on $n=10$ qubits with a depth of $L=30$. As shown here, different error models can be distinguished well, provided the output states are sufficiently far apart. Due to the large number of gates, in case the associated errors are large, the output state is almost maximally mixed. To demonstrate this, we additionally plot the entropy $S(\rho) = - \tr (\rho \log \rho)$ of the output state, $\rho$. Since all gates and errors in this example preserve the total parity $Z^{\otimes 10}$, we have $S(\rho) \leq 9$.}
    \label{fig:brickwall_10q_demo}
\end{figure*}

We sample $100$ random brickwall circuits (by the means of randomly choosing the individual gate parameters, see Fig.~\ref{fig:brickwall_layout}) on $n=8$ qubits of depth $L=24$, as well as $10$ circuits on $n=10$ qubits of depth $L=30$. Below, we also include an example of a circuit on $n=40$ qubits. The results show qualitatively (though with slight differences across the realizations) the same behaviour as in the previous examples: Given that the distance between two output states $\rho$ and $\sigma$ is sufficiently large, at least one of the statistical tests correctly detects the discrepancy in the errors. To demonstrate this, we depict in Fig.~\ref{fig:brickwall_10q_demo} the results for a specific realization on $n=10$ qubits. Note that in these circuits, the total number of gates is approximately $n(L-n)$, which is much larger than the roughly $2n$ gates in the previous examples. Due to this, the output state becomes close to maximally mixed in case the errors are large. This can also be seen in Fig.~\ref{fig:brickwall_10q_demo}, where we additionally plot the entropy of the output state. The output state corresponding to the error model we compare to (see the white cross) is, however, not completely mixed. Hence, in this example it is possible to distinguish the (highly noisy) output state from one which is completely mixed, as well as states arising from less noisy circuits.

\begin{figure}[htb!]
    \centering
    \includegraphics[width=\linewidth]{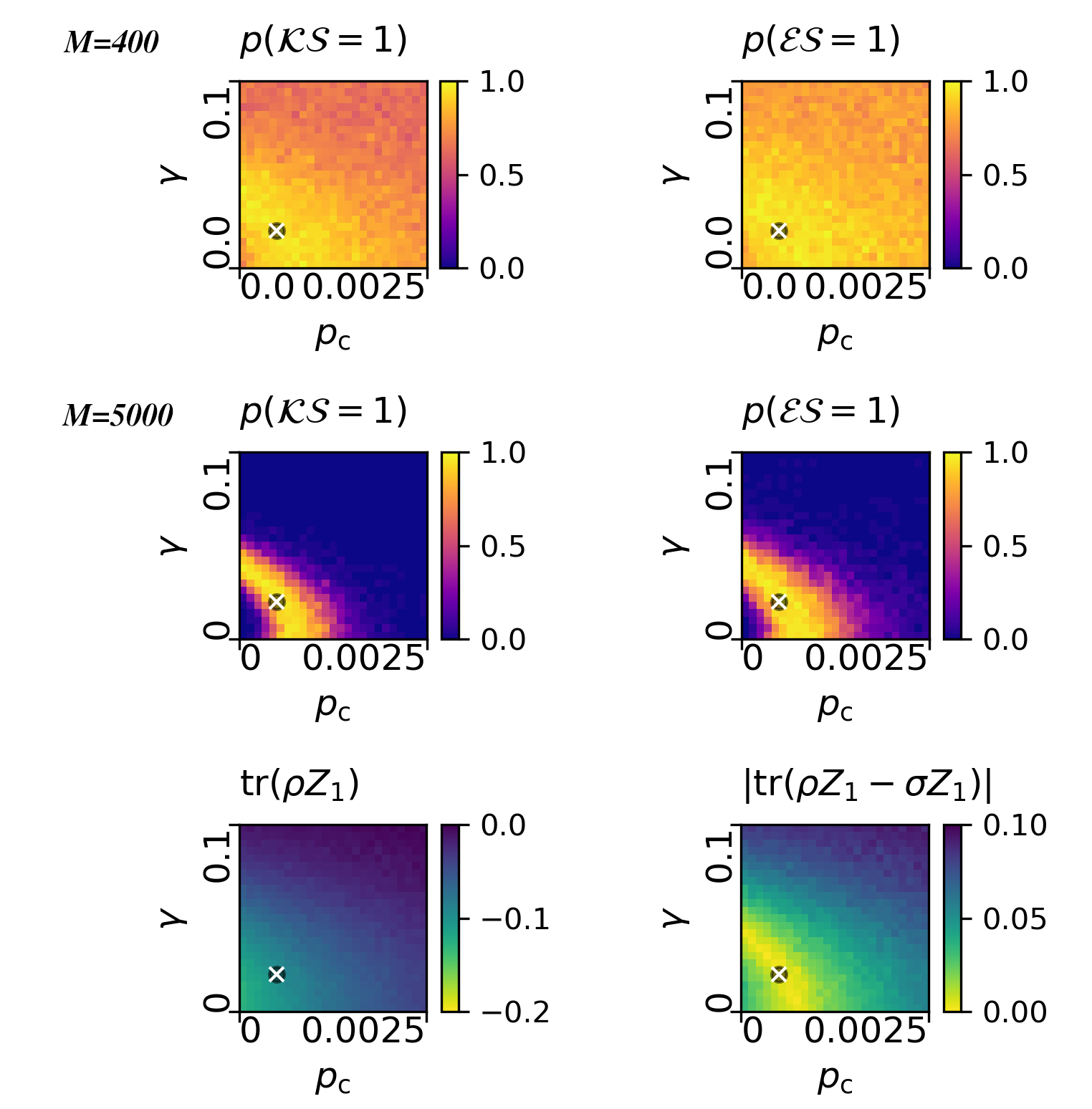}
    \caption{Comparing output states obtained from a random brickwall circuit on $n=40$ qubits with depth $L=120$ (see Fig.~\ref{fig:brickwall_layout} for details on the layout) using $M=400$ (top row) or $M=5000$ (middle row) measurement shots. It is evident from these plots that $M=400$ is (in contrast to $M=5000$) insufficient to estimate the errors. The plots show the same features as the other examples, however due to the large number of qubits we do not calculate anymore a distance in the output states $\rho$ and $\sigma$. To demonstrate that these plots reproduce the region in which the output states truly differ, we additionally plot differences in the expectation values of a simple observable. Note that over a large region, the errors are not large enough such that the output state is fully depolarized, as can be seen from the expectation value of $Z_1$.}
    \label{fig:brickwall_40q_demo}
\end{figure}

To demonstrate that our protocols are also practically scalable, we include an example of a brickwall circuit on $n=40$ qubits with a depth of $L=120$. By practically scalable we mean that the protocol is not only computationally viable in the sense that one can efficiently sample from the noisy output state, but is also capable of distinguishing output states corresponding to different error models on a large number of qubits. The values of the error parameters we have used for all the previous examples give rise to a completely mixed output state, since the system in question here is much larger compared to all the previous ones. To obtain output states which are not maximally mixed, we hence reduce the parameters $p_\mathrm{c}$ and $\gamma$. For instance, the error parameters of the reference state (white cross in Fig.~\ref{fig:brickwall_40q_demo}) are set to $p_c = 5 \cdot 10^{-4}$ and $\gamma = 0.02$.

As for such a large system, it is hard to calculate the full density matrices of the output states, we do not compute those and distances therein anymore. Instead, we resort to using the simulation techniques described in App.~\ref{appendix:simulation-of-mgs}. To get some confidence that a difference in the output states can be captured by the statistical tests, we estimate the quantity $\vert \tr (Z_1 \rho - Z_1 \sigma) \vert$ for the two output states $\rho$ and $\sigma$ from the obtained samples. To further demonstrate that the output state $\rho$ is not completely mixed for a wide range of error parameters, we also estimate the quantity $\tr (Z_1 \rho)$. Qualitatively, our example result (see Fig.~\ref{fig:brickwall_40q_demo}) reproduces all the features of the previous examples, namely that if there is a (significant) difference in the output states, the statistical tests correctly distinguish them. In this particular example, when using $M=400$ measurement shots one can already observe that the statistical tests start to distinguish different output states. However, for the range of parameters we consider, we do not obtain a better bound than $p(\mathcal(KS) = 1) < 0.5$ for vastly differing error models. It is hence necessary to increase the number of measurements. We find that using $M=5000$ gives satisfactory results (see Fig.~\ref{fig:brickwall_40q_demo}). Note that, as mentioned before, numerical simulations can be used to estimate the required number of measurements prior the quantum computation.

\subsection{Applying statistical tests to Haar-random and MG-Haar-random states} \label{appendix:haar-random-states}

Another application of the statistical tests we investigate here is how the tests perform in distinguishing random quantum states. We numerically analyze here two different notions of random states. On the one hand, we consider Haar-random states, and on the other hand, we consider states that arise from the action of random MG circuits (we refer to the latter notion as "MG-Haar-random"). Recall that MG circuits on $n$ qubits arise as a unitary representation of the orthogonal group acting on $\mathbb{R}^{2n}$ \cite{JozsaMiyake, TerhalMG}. To produce MG-Haar-random states, we first sample an orthogonal matrix according to the Haar measure on $\mathcal{SO}(2n)$, and then apply the corresponding MG circuit to $\ket{0^n}$\footnote{An alternative characterization of MG-generateable states is the set of all states $\ket \psi$, for which $Z^{\otimes n} \ket \psi = \ket \psi$ and $\Lambda \ket\psi^{\otimes 2} = 0$, where $\Lambda = \sum_{i=1}^{2n} c_i \otimes c_i$ and the $c_i$ are the Jordan-Wigner operators \cite{SpSc18}. Such states form a low-dimensional subset in the Hilbert space.}. We show here that 
MG-Haar-random states can be distinguished with high probability, which is, of course, in favor of our verification tests.

Let us now detail our setup: We sample $K = 5\times 10^4$ pairs of random states for each number of qubits $n\in\{7,\ldots,12\}$ in the Haar-random case and $n\in\{7,\ldots,16\}$ in the MG-Haar-random case. For each pair $(\ket \psi, \ket \phi)$, we estimate $p(\mathcal{KS}(\ket \psi, \ket \phi, M, \alpha) = 1)$ (similarly for the ES test) for various numbers of measurement shots $M$ and significance levels $\alpha$. The quantity we are interested in is the ratio $S/K$ of states that can be successfully distinguished. We choose here to say that two states can be  "successfully distinguished", if one can bound the probability that the test keeps the hypothesis "$\ket \phi = \ket \psi$", even though $\ket \psi \neq \ket \phi$, with the significance level $\alpha$. That is, we count the number of pairs for which
\[
p(\mathcal{KS}(\ket \psi, \ket \phi, M, \alpha) = 1 \big| \ket \psi \neq \ket \phi) \leq \alpha.
\]

\begin{figure*}[htb!]
    \centering
    \includegraphics[width=\linewidth]{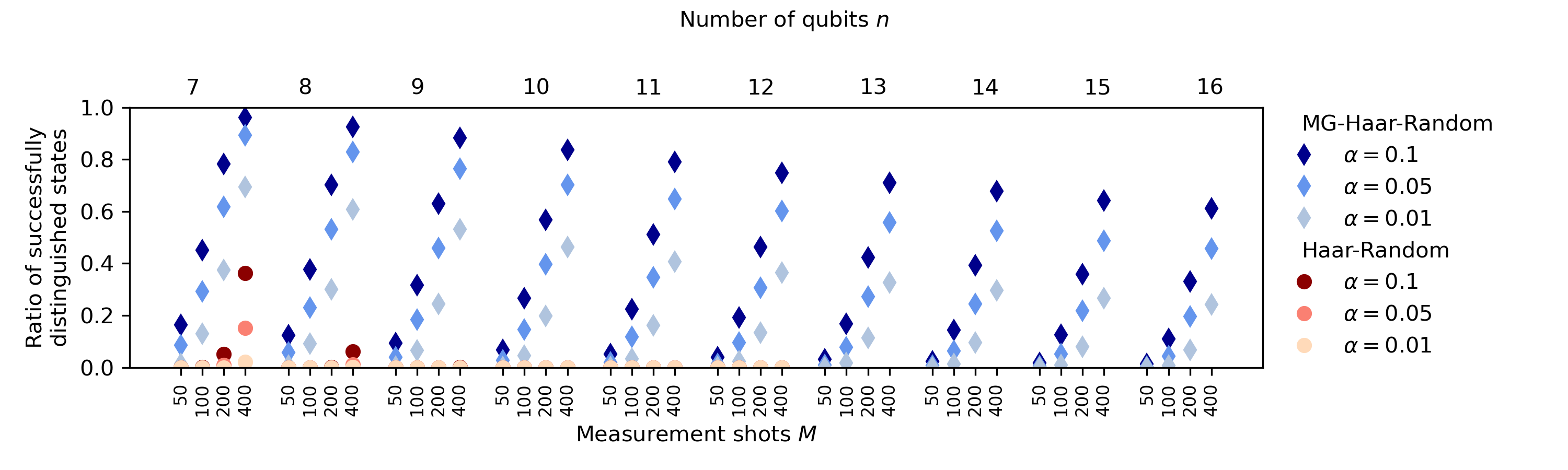}
    \caption{The ratio of successfully distinguished pairs of random states (see App.~\ref{appendix:haar-random-states}) obtained via the KS test. Pairs of states on various numbers of qubits $n$ are compared with different significance levels $\alpha$ and using different numbers of mesurement shots $M$. The MG-Haar-random states originate from the action of random MG circuits applied to the state $\ket{0^n}$. When using the ES test, we qualitatively observe the same behavior, albeit with slightly different ratios.}
    \label{fig:mg_haar_rand}
\end{figure*}

We choose here the significance level $\alpha$ to make it symmetric w.r.t. the other type of error that a statistical test can make: Rejecting the hypothesis "$\ket \psi = \ket \phi$" even though it is true, the probability of which can be bounded by $\alpha$. The expectation here is that with increasing $M$, the statistical tests should be able to distinguish two states better. Additionally, it is interesting to know how small $\alpha$ can be chosen such that a reasonably large fraction of states can still be distinguished. We find for our choices of the significance level $\alpha\in\{0.1,0.05,0.01\}$, a large fraction of MG-Haar-random states can be distinguished. More so, increasing this fraction can be done by increasing the number of measurement shots $M$ (here, we consider $M\in\{50,100,200,400\}$, see Fig.~\ref{fig:mg_haar_rand}). As can be seen in the figure, the ratio of successfully distinguished pairs descreases slightly with the number of qubits for $n=7,\ldots,16$. On the other hand, it appears that for our moderate choices of $n$ and $M$, Haar-random states cannot really be distinguished.

We remark that we also explicitly test the case where we generate two samples from the same state and apply the statistical test to it. Recall that the range of critical values for the test statistic is chosen in a way such that
\[
p(\mathcal{T}(\ket \psi, \ket \phi, M, \alpha) = 0 \big| \ket \psi = \ket \phi) \leq \alpha,
\]
for $\mathcal{T} = \mathcal{KS}, \mathcal{ES}$. In our simulations, we find that a better bound is given by $2\alpha$. A possible explanation for this discrepancy is that with the given choices of $M$, the limiting distribution of the test statistic is not yet reached (indeed, we find that for the ES test, as $M$ increases, so does the ratio of states for which $p(\mathcal{ES}=0) \leq \alpha$).


\begin{thebibliography}{37}%
\makeatletter
\providecommand \@ifxundefined [1]{%
 \@ifx{#1\undefined}
}%
\providecommand \@ifnum [1]{%
 \ifnum #1\expandafter \@firstoftwo
 \else \expandafter \@secondoftwo
 \fi
}%
\providecommand \@ifx [1]{%
 \ifx #1\expandafter \@firstoftwo
 \else \expandafter \@secondoftwo
 \fi
}%
\providecommand \natexlab [1]{#1}%
\providecommand \enquote  [1]{``#1''}%
\providecommand \bibnamefont  [1]{#1}%
\providecommand \bibfnamefont [1]{#1}%
\providecommand \citenamefont [1]{#1}%
\providecommand \href@noop [0]{\@secondoftwo}%
\providecommand \href [0]{\begingroup \@sanitize@url \@href}%
\providecommand \@href[1]{\@@startlink{#1}\@@href}%
\providecommand \@@href[1]{\endgroup#1\@@endlink}%
\providecommand \@sanitize@url [0]{\catcode `\\12\catcode `\$12\catcode
  `\&12\catcode `\#12\catcode `\^12\catcode `\_12\catcode `\%12\relax}%
\providecommand \@@startlink[1]{}%
\providecommand \@@endlink[0]{}%
\providecommand \url  [0]{\begingroup\@sanitize@url \@url }%
\providecommand \@url [1]{\endgroup\@href {#1}{\urlprefix }}%
\providecommand \urlprefix  [0]{URL }%
\providecommand \Eprint [0]{\href }%
\providecommand \doibase [0]{https://doi.org/}%
\providecommand \selectlanguage [0]{\@gobble}%
\providecommand \bibinfo  [0]{\@secondoftwo}%
\providecommand \bibfield  [0]{\@secondoftwo}%
\providecommand \translation [1]{[#1]}%
\providecommand \BibitemOpen [0]{}%
\providecommand \bibitemStop [0]{}%
\providecommand \bibitemNoStop [0]{.\EOS\space}%
\providecommand \EOS [0]{\spacefactor3000\relax}%
\providecommand \BibitemShut  [1]{\csname bibitem#1\endcsname}%
\let\auto@bib@innerbib\@empty
\bibitem [{\citenamefont {Mahadev}(2018)}]{Mah}%
  \BibitemOpen
  \bibfield  {author} {\bibinfo {author} {\bibfnamefont {U.}~\bibnamefont
  {Mahadev}},\ }\bibfield  {title} {\bibinfo {title} {Classical verification of
  quantum computations},\ }in\ \href {https://doi.org/10.1109/FOCS.2018.00033}
  {\emph {\bibinfo {booktitle} {{2018 IEEE 59th} {A}nnual {S}ymposium on
  {F}oundations of {C}omputer {S}cience ({FOCS})}}}\ (\bibinfo {year} {2018})\
  pp.\ \bibinfo {pages} {259--267}\BibitemShut {NoStop}%
\bibitem [{\citenamefont {Brakerski}\ \emph {et~al.}(2020)\citenamefont
  {Brakerski}, \citenamefont {Koppula}, \citenamefont {Vazirani},\ and\
  \citenamefont {Vidick}}]{Vidick1}%
  \BibitemOpen
  \bibfield  {author} {\bibinfo {author} {\bibfnamefont {Z.}~\bibnamefont
  {Brakerski}}, \bibinfo {author} {\bibfnamefont {V.}~\bibnamefont {Koppula}},
  \bibinfo {author} {\bibfnamefont {U.}~\bibnamefont {Vazirani}},\ and\
  \bibinfo {author} {\bibfnamefont {T.}~\bibnamefont {Vidick}},\ }\bibfield
  {title} {\bibinfo {title} {Simpler proofs of quantumness},\ }in\ \href
  {https://doi.org/10.4230/LIPIcs.TQC.2020.8} {\emph {\bibinfo {booktitle}
  {15th Conference on the Theory of Quantum Computation, Communication and
  Cryptography (TQC 2020)}}},\ \bibinfo {series} {Leibniz International
  Proceedings in Informatics (LIPIcs)}, Vol.\ \bibinfo {volume} {158},\
  \bibinfo {editor} {edited by\ \bibinfo {editor} {\bibfnamefont {S.~T.}\
  \bibnamefont {Flammia}}}\ (\bibinfo  {publisher} {Schloss
  Dagstuhl--Leibniz-Zentrum f{\"u}r Informatik},\ \bibinfo {address} {Dagstuhl,
  Germany},\ \bibinfo {year} {2020})\ pp.\ \bibinfo {pages}
  {8:1--8:14}\BibitemShut {NoStop}%
\bibitem [{\citenamefont {Brakerski}\ \emph {et~al.}(2021)\citenamefont
  {Brakerski}, \citenamefont {Christiano}, \citenamefont {Mahadev},
  \citenamefont {Vazirani},\ and\ \citenamefont {Vidick}}]{Vidick2}%
  \BibitemOpen
  \bibfield  {author} {\bibinfo {author} {\bibfnamefont {Z.}~\bibnamefont
  {Brakerski}}, \bibinfo {author} {\bibfnamefont {P.}~\bibnamefont
  {Christiano}}, \bibinfo {author} {\bibfnamefont {U.}~\bibnamefont {Mahadev}},
  \bibinfo {author} {\bibfnamefont {U.}~\bibnamefont {Vazirani}},\ and\
  \bibinfo {author} {\bibfnamefont {T.}~\bibnamefont {Vidick}},\ }\bibfield
  {title} {\bibinfo {title} {A cryptographic test of quantumness and
  certifiable randomness from a single quantum device},\ }\href
  {https://doi.org/10.1145/3441309} {\bibfield  {journal} {\bibinfo  {journal}
  {Journal of the ACM}\ }\textbf {\bibinfo {volume} {68}},\ \bibinfo {pages}
  {1} (\bibinfo {year} {2021})}\BibitemShut {NoStop}%
\bibitem [{\citenamefont {Eisert}\ \emph {et~al.}(2020)\citenamefont {Eisert},
  \citenamefont {Hangleiter}, \citenamefont {Walk}, \citenamefont {Roth},
  \citenamefont {Markham}, \citenamefont {Parekh}, \citenamefont {Chabaud},\
  and\ \citenamefont {Kashefi}}]{Ei20}%
  \BibitemOpen
  \bibfield  {author} {\bibinfo {author} {\bibfnamefont {J.}~\bibnamefont
  {Eisert}}, \bibinfo {author} {\bibfnamefont {D.}~\bibnamefont {Hangleiter}},
  \bibinfo {author} {\bibfnamefont {N.}~\bibnamefont {Walk}}, \bibinfo {author}
  {\bibfnamefont {I.}~\bibnamefont {Roth}}, \bibinfo {author} {\bibfnamefont
  {D.}~\bibnamefont {Markham}}, \bibinfo {author} {\bibfnamefont
  {R.}~\bibnamefont {Parekh}}, \bibinfo {author} {\bibfnamefont
  {U.}~\bibnamefont {Chabaud}},\ and\ \bibinfo {author} {\bibfnamefont
  {E.}~\bibnamefont {Kashefi}},\ }\bibfield  {title} {\bibinfo {title} {Quantum
  certification and benchmarking},\ }\href
  {https://doi.org/10.1038/s42254-020-0186-4} {\bibfield  {journal} {\bibinfo
  {journal} {Nature Reviews Physics}\ }\textbf {\bibinfo {volume} {2}},\
  \bibinfo {pages} {382} (\bibinfo {year} {2020})}\BibitemShut {NoStop}%
\bibitem [{\citenamefont {Knill}\ \emph {et~al.}(2008)\citenamefont {Knill},
  \citenamefont {Leibfried}, \citenamefont {Reichle}, \citenamefont {Britton},
  \citenamefont {Blakestad}, \citenamefont {Jost}, \citenamefont {Langer},
  \citenamefont {Ozeri}, \citenamefont {Seidelin},\ and\ \citenamefont
  {Wineland}}]{KnillRB}%
  \BibitemOpen
  \bibfield  {author} {\bibinfo {author} {\bibfnamefont {E.}~\bibnamefont
  {Knill}}, \bibinfo {author} {\bibfnamefont {D.}~\bibnamefont {Leibfried}},
  \bibinfo {author} {\bibfnamefont {R.}~\bibnamefont {Reichle}}, \bibinfo
  {author} {\bibfnamefont {J.}~\bibnamefont {Britton}}, \bibinfo {author}
  {\bibfnamefont {R.~B.}\ \bibnamefont {Blakestad}}, \bibinfo {author}
  {\bibfnamefont {J.~D.}\ \bibnamefont {Jost}}, \bibinfo {author}
  {\bibfnamefont {C.}~\bibnamefont {Langer}}, \bibinfo {author} {\bibfnamefont
  {R.}~\bibnamefont {Ozeri}}, \bibinfo {author} {\bibfnamefont
  {S.}~\bibnamefont {Seidelin}},\ and\ \bibinfo {author} {\bibfnamefont
  {D.~J.}\ \bibnamefont {Wineland}},\ }\bibfield  {title} {\bibinfo {title}
  {Randomized benchmarking of quantum gates},\ }\href
  {https://doi.org/10.1103/PhysRevA.77.012307} {\bibfield  {journal} {\bibinfo
  {journal} {Phys. Rev. A}\ }\textbf {\bibinfo {volume} {77}},\ \bibinfo
  {pages} {012307} (\bibinfo {year} {2008})}\BibitemShut {NoStop}%
\bibitem [{\citenamefont {Lu}\ \emph {et~al.}(2015)\citenamefont {Lu},
  \citenamefont {Li}, \citenamefont {Trottier}, \citenamefont {Li},
  \citenamefont {Brodutch}, \citenamefont {Krismanich}, \citenamefont
  {Ghavami}, \citenamefont {Dmitrienko}, \citenamefont {Long}, \citenamefont
  {Baugh},\ and\ \citenamefont {Laflamme}}]{LuLi15}%
  \BibitemOpen
  \bibfield  {author} {\bibinfo {author} {\bibfnamefont {D.}~\bibnamefont
  {Lu}}, \bibinfo {author} {\bibfnamefont {H.}~\bibnamefont {Li}}, \bibinfo
  {author} {\bibfnamefont {D.-A.}\ \bibnamefont {Trottier}}, \bibinfo {author}
  {\bibfnamefont {J.}~\bibnamefont {Li}}, \bibinfo {author} {\bibfnamefont
  {A.}~\bibnamefont {Brodutch}}, \bibinfo {author} {\bibfnamefont {A.~P.}\
  \bibnamefont {Krismanich}}, \bibinfo {author} {\bibfnamefont
  {A.}~\bibnamefont {Ghavami}}, \bibinfo {author} {\bibfnamefont {G.~I.}\
  \bibnamefont {Dmitrienko}}, \bibinfo {author} {\bibfnamefont
  {G.}~\bibnamefont {Long}}, \bibinfo {author} {\bibfnamefont {J.}~\bibnamefont
  {Baugh}},\ and\ \bibinfo {author} {\bibfnamefont {R.}~\bibnamefont
  {Laflamme}},\ }\bibfield  {title} {\bibinfo {title} {Experimental estimation
  of average fidelity of a clifford gate on a 7-qubit quantum processor},\
  }\href {https://doi.org/10.1103/PhysRevLett.114.140505} {\bibfield  {journal}
  {\bibinfo  {journal} {Phys. Rev. Lett.}\ }\textbf {\bibinfo {volume} {114}},\
  \bibinfo {pages} {140505} (\bibinfo {year} {2015})}\BibitemShut {NoStop}%
\bibitem [{\citenamefont {Magesan}\ \emph {et~al.}(2012)\citenamefont
  {Magesan}, \citenamefont {Gambetta}, \citenamefont {Johnson}, \citenamefont
  {Ryan}, \citenamefont {Chow}, \citenamefont {Merkel}, \citenamefont
  {da~Silva}, \citenamefont {Keefe}, \citenamefont {Rothwell}, \citenamefont
  {Ohki}, \citenamefont {Ketchen},\ and\ \citenamefont {Steffen}}]{MaGa12}%
  \BibitemOpen
  \bibfield  {author} {\bibinfo {author} {\bibfnamefont {E.}~\bibnamefont
  {Magesan}}, \bibinfo {author} {\bibfnamefont {J.~M.}\ \bibnamefont
  {Gambetta}}, \bibinfo {author} {\bibfnamefont {B.~R.}\ \bibnamefont
  {Johnson}}, \bibinfo {author} {\bibfnamefont {C.~A.}\ \bibnamefont {Ryan}},
  \bibinfo {author} {\bibfnamefont {J.~M.}\ \bibnamefont {Chow}}, \bibinfo
  {author} {\bibfnamefont {S.~T.}\ \bibnamefont {Merkel}}, \bibinfo {author}
  {\bibfnamefont {M.~P.}\ \bibnamefont {da~Silva}}, \bibinfo {author}
  {\bibfnamefont {G.~A.}\ \bibnamefont {Keefe}}, \bibinfo {author}
  {\bibfnamefont {M.~B.}\ \bibnamefont {Rothwell}}, \bibinfo {author}
  {\bibfnamefont {T.~A.}\ \bibnamefont {Ohki}}, \bibinfo {author}
  {\bibfnamefont {M.~B.}\ \bibnamefont {Ketchen}},\ and\ \bibinfo {author}
  {\bibfnamefont {M.}~\bibnamefont {Steffen}},\ }\bibfield  {title} {\bibinfo
  {title} {Efficient measurement of quantum gate error by interleaved
  randomized benchmarking},\ }\href
  {https://doi.org/10.1103/PhysRevLett.109.080505} {\bibfield  {journal}
  {\bibinfo  {journal} {Phys. Rev. Lett.}\ }\textbf {\bibinfo {volume} {109}},\
  \bibinfo {pages} {080505} (\bibinfo {year} {2012})}\BibitemShut {NoStop}%
\bibitem [{\citenamefont {Onorati}\ \emph {et~al.}(2019)\citenamefont
  {Onorati}, \citenamefont {Werner},\ and\ \citenamefont {Eisert}}]{OnWe19}%
  \BibitemOpen
  \bibfield  {author} {\bibinfo {author} {\bibfnamefont {E.}~\bibnamefont
  {Onorati}}, \bibinfo {author} {\bibfnamefont {A.~H.}\ \bibnamefont
  {Werner}},\ and\ \bibinfo {author} {\bibfnamefont {J.}~\bibnamefont
  {Eisert}},\ }\bibfield  {title} {\bibinfo {title} {Randomized benchmarking
  for individual quantum gates},\ }\href
  {https://doi.org/10.1103/PhysRevLett.123.060501} {\bibfield  {journal}
  {\bibinfo  {journal} {Phys. Rev. Lett.}\ }\textbf {\bibinfo {volume} {123}},\
  \bibinfo {pages} {060501} (\bibinfo {year} {2019})}\BibitemShut {NoStop}%
\bibitem [{\citenamefont {Proctor}\ \emph
  {et~al.}(2022{\natexlab{a}})\citenamefont {Proctor}, \citenamefont {Seritan},
  \citenamefont {Rudinger}, \citenamefont {Nielsen}, \citenamefont
  {Blume-Kohout},\ and\ \citenamefont {Young}}]{MirrorRB}%
  \BibitemOpen
  \bibfield  {author} {\bibinfo {author} {\bibfnamefont {T.}~\bibnamefont
  {Proctor}}, \bibinfo {author} {\bibfnamefont {S.}~\bibnamefont {Seritan}},
  \bibinfo {author} {\bibfnamefont {K.}~\bibnamefont {Rudinger}}, \bibinfo
  {author} {\bibfnamefont {E.}~\bibnamefont {Nielsen}}, \bibinfo {author}
  {\bibfnamefont {R.}~\bibnamefont {Blume-Kohout}},\ and\ \bibinfo {author}
  {\bibfnamefont {K.}~\bibnamefont {Young}},\ }\bibfield  {title} {\bibinfo
  {title} {Scalable randomized benchmarking of quantum computers using mirror
  circuits},\ }\href {https://doi.org/10.1103/PhysRevLett.129.150502}
  {\bibfield  {journal} {\bibinfo  {journal} {Phys. Rev. Lett.}\ }\textbf
  {\bibinfo {volume} {129}},\ \bibinfo {pages} {150502} (\bibinfo {year}
  {2022}{\natexlab{a}})}\BibitemShut {NoStop}%
\bibitem [{\citenamefont {Nielsen}\ and\ \citenamefont {Chuang}(2010)}]{NC10}%
  \BibitemOpen
  \bibfield  {author} {\bibinfo {author} {\bibfnamefont {M.~A.}\ \bibnamefont
  {Nielsen}}\ and\ \bibinfo {author} {\bibfnamefont {I.~L.}\ \bibnamefont
  {Chuang}},\ }\href@noop {} {\emph {\bibinfo {title} {Quantum Computation and
  Quantum Information}}},\ \bibinfo {edition} {10th}\ ed.\ (\bibinfo
  {publisher} {Cambridge University Press},\ \bibinfo {address} {Cambridge},\
  \bibinfo {year} {2010})\BibitemShut {NoStop}%
\bibitem [{\citenamefont {Jozsa}\ and\ \citenamefont {Strelchuk}(2017)}]{JS17}%
  \BibitemOpen
  \bibfield  {author} {\bibinfo {author} {\bibfnamefont {R.}~\bibnamefont
  {Jozsa}}\ and\ \bibinfo {author} {\bibfnamefont {S.}~\bibnamefont
  {Strelchuk}},\ }\href@noop {} {\bibinfo {title} {Efficient classical
  verification of quantum computations}} (\bibinfo {year} {2017}),\ \Eprint
  {https://arxiv.org/abs/1705.02817} {arXiv:1705.02817 [quant-ph]} \BibitemShut
  {NoStop}%
\bibitem [{\citenamefont {Ferracin}\ \emph {et~al.}(2019)\citenamefont
  {Ferracin}, \citenamefont {Kapourniotis},\ and\ \citenamefont
  {Datta}}]{FeKa19}%
  \BibitemOpen
  \bibfield  {author} {\bibinfo {author} {\bibfnamefont {S.}~\bibnamefont
  {Ferracin}}, \bibinfo {author} {\bibfnamefont {T.}~\bibnamefont
  {Kapourniotis}},\ and\ \bibinfo {author} {\bibfnamefont {A.}~\bibnamefont
  {Datta}},\ }\bibfield  {title} {\bibinfo {title} {Accrediting outputs of
  noisy intermediate-scale quantum computing devices},\ }\href
  {https://doi.org/10.1088/1367-2630/ab4fd6} {\bibfield  {journal} {\bibinfo
  {journal} {New. J. Phys.}\ }\textbf {\bibinfo {volume} {21}},\ \bibinfo
  {pages} {113038} (\bibinfo {year} {2019})}\BibitemShut {NoStop}%
\bibitem [{\citenamefont {Schuch}\ \emph {et~al.}(2008)\citenamefont {Schuch},
  \citenamefont {Wolf}, \citenamefont {Vollbrecht},\ and\ \citenamefont
  {I.}}]{ScCi08}%
  \BibitemOpen
  \bibfield  {author} {\bibinfo {author} {\bibfnamefont {N.}~\bibnamefont
  {Schuch}}, \bibinfo {author} {\bibfnamefont {M.~M.}\ \bibnamefont {Wolf}},
  \bibinfo {author} {\bibfnamefont {K.~G.~H.}\ \bibnamefont {Vollbrecht}},\
  and\ \bibinfo {author} {\bibfnamefont {C.~J.}\ \bibnamefont {I.}},\
  }\bibfield  {title} {\bibinfo {title} {On entropy growth and the hardness of
  simulating time evolution},\ }\href
  {https://doi.org/10.1088/1367-2630/10/3/033032} {\bibfield  {journal}
  {\bibinfo  {journal} {New. J. Phys.}\ }\textbf {\bibinfo {volume} {10}},\
  \bibinfo {pages} {033032} (\bibinfo {year} {2008})}\BibitemShut {NoStop}%
\bibitem [{\citenamefont {Elben}\ \emph {et~al.}(2020)\citenamefont {Elben},
  \citenamefont {Vermersch}, \citenamefont {van Bijnen}, \citenamefont
  {Kokail}, \citenamefont {Brydges}, \citenamefont {Maier}, \citenamefont
  {Joshi}, \citenamefont {Blatt}, \citenamefont {Roos},\ and\ \citenamefont
  {Zoller}}]{ElbenZoller}%
  \BibitemOpen
  \bibfield  {author} {\bibinfo {author} {\bibfnamefont {A.}~\bibnamefont
  {Elben}}, \bibinfo {author} {\bibfnamefont {B.}~\bibnamefont {Vermersch}},
  \bibinfo {author} {\bibfnamefont {R.}~\bibnamefont {van Bijnen}}, \bibinfo
  {author} {\bibfnamefont {C.}~\bibnamefont {Kokail}}, \bibinfo {author}
  {\bibfnamefont {T.}~\bibnamefont {Brydges}}, \bibinfo {author} {\bibfnamefont
  {C.}~\bibnamefont {Maier}}, \bibinfo {author} {\bibfnamefont {M.~K.}\
  \bibnamefont {Joshi}}, \bibinfo {author} {\bibfnamefont {R.}~\bibnamefont
  {Blatt}}, \bibinfo {author} {\bibfnamefont {C.~F.}\ \bibnamefont {Roos}},\
  and\ \bibinfo {author} {\bibfnamefont {P.}~\bibnamefont {Zoller}},\
  }\bibfield  {title} {\bibinfo {title} {Cross-platform verification of
  intermediate scale quantum devices},\ }\href
  {https://doi.org/10.1103/PhysRevLett.124.010504} {\bibfield  {journal}
  {\bibinfo  {journal} {Phys. Rev. Lett.}\ }\textbf {\bibinfo {volume} {124}},\
  \bibinfo {pages} {010504} (\bibinfo {year} {2020})}\BibitemShut {NoStop}%
\bibitem [{\citenamefont {Hebenstreit}\ \emph {et~al.}(2019)\citenamefont
  {Hebenstreit}, \citenamefont {Jozsa}, \citenamefont {Kraus}, \citenamefont
  {Strelchuk},\ and\ \citenamefont {Yoganathan}}]{Hebenstreit2019}%
  \BibitemOpen
  \bibfield  {author} {\bibinfo {author} {\bibfnamefont {M.}~\bibnamefont
  {Hebenstreit}}, \bibinfo {author} {\bibfnamefont {R.}~\bibnamefont {Jozsa}},
  \bibinfo {author} {\bibfnamefont {B.}~\bibnamefont {Kraus}}, \bibinfo
  {author} {\bibfnamefont {S.}~\bibnamefont {Strelchuk}},\ and\ \bibinfo
  {author} {\bibfnamefont {M.}~\bibnamefont {Yoganathan}},\ }\bibfield  {title}
  {\bibinfo {title} {All pure fermionic non-{Gaussian} states are magic states
  for matchgate computations},\ }\href
  {https://doi.org/10.1103/PhysRevLett.123.080503} {\bibfield  {journal}
  {\bibinfo  {journal} {Phys. Rev. Lett.}\ }\textbf {\bibinfo {volume} {123}},\
  \bibinfo {pages} {080503} (\bibinfo {year} {2019})}\BibitemShut {NoStop}%
\bibitem [{\citenamefont {Van Den~Nest}(2011)}]{Ne11}%
  \BibitemOpen
  \bibfield  {author} {\bibinfo {author} {\bibfnamefont {M.}~\bibnamefont {Van
  Den~Nest}},\ }\bibfield  {title} {\bibinfo {title} {Simulating quantum
  computers with probabilistic methods},\ }\href
  {https://dl.acm.org/doi/10.5555/2230936.2230941} {\bibfield  {journal}
  {\bibinfo  {journal} {Quantum Info. Comput.}\ }\textbf {\bibinfo {volume}
  {11}},\ \bibinfo {pages} {784–812} (\bibinfo {year} {2011})}\BibitemShut
  {NoStop}%
\bibitem [{\citenamefont {Wallman}\ and\ \citenamefont
  {Emerson}(2016)}]{EmersonRandComp}%
  \BibitemOpen
  \bibfield  {author} {\bibinfo {author} {\bibfnamefont {J.~J.}\ \bibnamefont
  {Wallman}}\ and\ \bibinfo {author} {\bibfnamefont {J.}~\bibnamefont
  {Emerson}},\ }\bibfield  {title} {\bibinfo {title} {Noise tailoring for
  scalable quantum computation via randomized compiling},\ }\href
  {https://doi.org/10.1103/PhysRevA.94.052325} {\bibfield  {journal} {\bibinfo
  {journal} {Phys. Rev. A}\ }\textbf {\bibinfo {volume} {94}},\ \bibinfo
  {pages} {052325} (\bibinfo {year} {2016})}\BibitemShut {NoStop}%
\bibitem [{\citenamefont {Valiant}(2002)}]{Valiant}%
  \BibitemOpen
  \bibfield  {author} {\bibinfo {author} {\bibfnamefont {L.~G.}\ \bibnamefont
  {Valiant}},\ }\bibfield  {title} {\bibinfo {title} {Quantum circuits that can
  be simulated classically in polynomial time},\ }\href
  {https://doi.org/10.1137/S0097539700377025} {\bibfield  {journal} {\bibinfo
  {journal} {SIAM Journal on Computing}\ }\textbf {\bibinfo {volume} {31}},\
  \bibinfo {pages} {1229} (\bibinfo {year} {2002})}\BibitemShut {NoStop}%
\bibitem [{\citenamefont {Terhal}\ and\ \citenamefont
  {DiVincenzo}(2002)}]{TerhalMG}%
  \BibitemOpen
  \bibfield  {author} {\bibinfo {author} {\bibfnamefont {B.~M.}\ \bibnamefont
  {Terhal}}\ and\ \bibinfo {author} {\bibfnamefont {D.~P.}\ \bibnamefont
  {DiVincenzo}},\ }\bibfield  {title} {\bibinfo {title} {Classical simulation
  of noninteracting-fermion quantum circuits},\ }\href
  {https://doi.org/10.1103/PhysRevA.65.032325} {\bibfield  {journal} {\bibinfo
  {journal} {Phys. Rev. A}\ }\textbf {\bibinfo {volume} {65}},\ \bibinfo
  {pages} {032325} (\bibinfo {year} {2002})}\BibitemShut {NoStop}%
\bibitem [{\citenamefont {Jozsa}\ and\ \citenamefont
  {Miyake}(2008)}]{JozsaMiyake}%
  \BibitemOpen
  \bibfield  {author} {\bibinfo {author} {\bibfnamefont {R.}~\bibnamefont
  {Jozsa}}\ and\ \bibinfo {author} {\bibfnamefont {A.}~\bibnamefont {Miyake}},\
  }\bibfield  {title} {\bibinfo {title} {Matchgates and classical simulation of
  quantum circuits},\ }\href {https://doi.org/10.1098/rspa.2008.0189}
  {\bibfield  {journal} {\bibinfo  {journal} {Proc. R. Soc. A}\ }\textbf
  {\bibinfo {volume} {464}},\ \bibinfo {pages} {3089} (\bibinfo {year}
  {2008})}\BibitemShut {NoStop}%
\bibitem [{\citenamefont {Jozsa}\ \emph {et~al.}(2010)\citenamefont {Jozsa},
  \citenamefont {Kraus}, \citenamefont {Miyake},\ and\ \citenamefont
  {Watrous}}]{JoszaMiyakeBKWotrous}%
  \BibitemOpen
  \bibfield  {author} {\bibinfo {author} {\bibfnamefont {R.}~\bibnamefont
  {Jozsa}}, \bibinfo {author} {\bibfnamefont {B.}~\bibnamefont {Kraus}},
  \bibinfo {author} {\bibfnamefont {A.}~\bibnamefont {Miyake}},\ and\ \bibinfo
  {author} {\bibfnamefont {J.}~\bibnamefont {Watrous}},\ }\bibfield  {title}
  {\bibinfo {title} {Matchgate and space-bounded quantum computations are
  equivalent},\ }\href {https://doi.org/10.1098/rspa.2009.0433} {\bibfield
  {journal} {\bibinfo  {journal} {Proc. R. Soc. A}\ }\textbf {\bibinfo {volume}
  {466}},\ \bibinfo {pages} {809} (\bibinfo {year} {2010})}\BibitemShut
  {NoStop}%
\bibitem [{\citenamefont {Brod}(2016)}]{Br16}%
  \BibitemOpen
  \bibfield  {author} {\bibinfo {author} {\bibfnamefont {D.~J.}\ \bibnamefont
  {Brod}},\ }\bibfield  {title} {\bibinfo {title} {Efficient classical
  simulation of matchgate circuits with generalized inputs and measurements},\
  }\href {https://doi.org/10.1103/PhysRevA.93.062332} {\bibfield  {journal}
  {\bibinfo  {journal} {Phys. Rev. A}\ }\textbf {\bibinfo {volume} {93}},\
  \bibinfo {pages} {062332} (\bibinfo {year} {2016})}\BibitemShut {NoStop}%
\bibitem [{\citenamefont {Hebenstreit}\ \emph {et~al.}(2020)\citenamefont
  {Hebenstreit}, \citenamefont {Jozsa}, \citenamefont {Kraus},\ and\
  \citenamefont {Strelchuk}}]{Hebenstreit2020}%
  \BibitemOpen
  \bibfield  {author} {\bibinfo {author} {\bibfnamefont {M.}~\bibnamefont
  {Hebenstreit}}, \bibinfo {author} {\bibfnamefont {R.}~\bibnamefont {Jozsa}},
  \bibinfo {author} {\bibfnamefont {B.}~\bibnamefont {Kraus}},\ and\ \bibinfo
  {author} {\bibfnamefont {S.}~\bibnamefont {Strelchuk}},\ }\bibfield  {title}
  {\bibinfo {title} {Computational power of matchgates with supplementary
  resources},\ }\href {https://doi.org/10.1103/PhysRevA.102.052604} {\bibfield
  {journal} {\bibinfo  {journal} {Phys. Rev. A}\ }\textbf {\bibinfo {volume}
  {102}},\ \bibinfo {pages} {052604} (\bibinfo {year} {2020})}\BibitemShut
  {NoStop}%
\bibitem [{\citenamefont {Brod}\ and\ \citenamefont
  {Galv\~ao}(2011)}]{Brod2011ExtendingMGs}%
  \BibitemOpen
  \bibfield  {author} {\bibinfo {author} {\bibfnamefont {D.~J.}\ \bibnamefont
  {Brod}}\ and\ \bibinfo {author} {\bibfnamefont {E.~F.}\ \bibnamefont
  {Galv\~ao}},\ }\bibfield  {title} {\bibinfo {title} {Extending matchgates
  into universal quantum computation},\ }\href
  {https://doi.org/10.1103/PhysRevA.84.022310} {\bibfield  {journal} {\bibinfo
  {journal} {Phys. Rev. A}\ }\textbf {\bibinfo {volume} {84}},\ \bibinfo
  {pages} {022310} (\bibinfo {year} {2011})}\BibitemShut {NoStop}%
\bibitem [{\citenamefont {Kolmogorov}(1933)}]{Ko33}%
  \BibitemOpen
  \bibfield  {author} {\bibinfo {author} {\bibfnamefont {A.}~\bibnamefont
  {Kolmogorov}},\ }\bibfield  {title} {\bibinfo {title} {Sulla determinazione
  empirica di una legge di distributione},\ }\href@noop {} {\bibfield
  {journal} {\bibinfo  {journal} {Giornale dell’ Istituto Italiano degli
  Attuari}\ }\textbf {\bibinfo {volume} {4}},\ \bibinfo {pages} {83} (\bibinfo
  {year} {1933})}\BibitemShut {NoStop}%
\bibitem [{\citenamefont {Epps}\ and\ \citenamefont
  {Singleton}(1986)}]{EppsSingleton}%
  \BibitemOpen
  \bibfield  {author} {\bibinfo {author} {\bibfnamefont {T.}~\bibnamefont
  {Epps}}\ and\ \bibinfo {author} {\bibfnamefont {K.~J.}\ \bibnamefont
  {Singleton}},\ }\bibfield  {title} {\bibinfo {title} {An omnibus test for the
  two-sample problem using the empirical characteristic function},\ }\href
  {https://doi.org/10.1080/00949658608810963} {\bibfield  {journal} {\bibinfo
  {journal} {Journal of Statistical Computation and Simulation}\ }\textbf
  {\bibinfo {volume} {26}},\ \bibinfo {pages} {177} (\bibinfo {year}
  {1986})}\BibitemShut {NoStop}%
\bibitem [{\citenamefont {Proctor}\ \emph
  {et~al.}(2022{\natexlab{b}})\citenamefont {Proctor}, \citenamefont {Seritan},
  \citenamefont {Nielsen}, \citenamefont {Rudinger}, \citenamefont {Young},
  \citenamefont {Blume-Kohout},\ and\ \citenamefont
  {Sarovar}}]{MirrorCircuitsTrust}%
  \BibitemOpen
  \bibfield  {author} {\bibinfo {author} {\bibfnamefont {T.}~\bibnamefont
  {Proctor}}, \bibinfo {author} {\bibfnamefont {S.}~\bibnamefont {Seritan}},
  \bibinfo {author} {\bibfnamefont {E.}~\bibnamefont {Nielsen}}, \bibinfo
  {author} {\bibfnamefont {K.}~\bibnamefont {Rudinger}}, \bibinfo {author}
  {\bibfnamefont {K.}~\bibnamefont {Young}}, \bibinfo {author} {\bibfnamefont
  {R.}~\bibnamefont {Blume-Kohout}},\ and\ \bibinfo {author} {\bibfnamefont
  {M.}~\bibnamefont {Sarovar}},\ }\href@noop {} {\bibinfo {title} {Establishing
  trust in quantum computations}} (\bibinfo {year} {2022}{\natexlab{b}}),\
  \Eprint {https://arxiv.org/abs/2204.07568} {arXiv:2204.07568 [quant-ph]}
  \BibitemShut {NoStop}%
\bibitem [{\citenamefont {Soeda}\ \emph {et~al.}(2014)\citenamefont {Soeda},
  \citenamefont {Akibue},\ and\ \citenamefont {Murao}}]{Murao}%
  \BibitemOpen
  \bibfield  {author} {\bibinfo {author} {\bibfnamefont {A.}~\bibnamefont
  {Soeda}}, \bibinfo {author} {\bibfnamefont {S.}~\bibnamefont {Akibue}},\ and\
  \bibinfo {author} {\bibfnamefont {M.}~\bibnamefont {Murao}},\ }\bibfield
  {title} {\bibinfo {title} {Two-party locc convertibility of quadpartite
  states and {Kraus–Cirac} number of two-qubit unitaries},\ }\href
  {https://doi.org/10.1088/1751-8113/47/42/424036} {\bibfield  {journal}
  {\bibinfo  {journal} {J. Phys. A: Math. Theor.}\ }\textbf {\bibinfo {volume}
  {47}},\ \bibinfo {pages} {424036} (\bibinfo {year} {2014})}\BibitemShut
  {NoStop}%
\bibitem [{\citenamefont {Heu\ss{}en}\ \emph {et~al.}(2023)\citenamefont
  {Heu\ss{}en}, \citenamefont {Postler}, \citenamefont {Rispler}, \citenamefont
  {Pogorelov}, \citenamefont {Marciniak}, \citenamefont {Monz}, \citenamefont
  {Schindler},\ and\ \citenamefont {M\"uller}}]{HeussenIonErrorModel}%
  \BibitemOpen
  \bibfield  {author} {\bibinfo {author} {\bibfnamefont {S.}~\bibnamefont
  {Heu\ss{}en}}, \bibinfo {author} {\bibfnamefont {L.}~\bibnamefont {Postler}},
  \bibinfo {author} {\bibfnamefont {M.}~\bibnamefont {Rispler}}, \bibinfo
  {author} {\bibfnamefont {I.}~\bibnamefont {Pogorelov}}, \bibinfo {author}
  {\bibfnamefont {C.~D.}\ \bibnamefont {Marciniak}}, \bibinfo {author}
  {\bibfnamefont {T.}~\bibnamefont {Monz}}, \bibinfo {author} {\bibfnamefont
  {P.}~\bibnamefont {Schindler}},\ and\ \bibinfo {author} {\bibfnamefont
  {M.}~\bibnamefont {M\"uller}},\ }\bibfield  {title} {\bibinfo {title}
  {Strategies for a practical advantage of fault-tolerant circuit design in
  noisy trapped-ion quantum computers},\ }\href
  {https://doi.org/10.1103/PhysRevA.107.042422} {\bibfield  {journal} {\bibinfo
   {journal} {Phys. Rev. A}\ }\textbf {\bibinfo {volume} {107}},\ \bibinfo
  {pages} {042422} (\bibinfo {year} {2023})}\BibitemShut {NoStop}%
\bibitem [{\citenamefont {Montanaro}\ and\ \citenamefont
  {Stanisic}(2021)}]{FermionicErrorMitigation}%
  \BibitemOpen
  \bibfield  {author} {\bibinfo {author} {\bibfnamefont {A.}~\bibnamefont
  {Montanaro}}\ and\ \bibinfo {author} {\bibfnamefont {S.}~\bibnamefont
  {Stanisic}},\ }\href@noop {} {\bibinfo {title} {Error mitigation by training
  with fermionic linear optics}} (\bibinfo {year} {2021}),\ \Eprint
  {https://arxiv.org/abs/2102.02120} {arXiv:2102.02120 [quant-ph]} \BibitemShut
  {NoStop}%
\bibitem [{\citenamefont {Jozsa}\ and\ \citenamefont {Van
  Den~Nest}(2014)}]{MaartenJozsa}%
  \BibitemOpen
  \bibfield  {author} {\bibinfo {author} {\bibfnamefont {R.}~\bibnamefont
  {Jozsa}}\ and\ \bibinfo {author} {\bibfnamefont {M.}~\bibnamefont {Van
  Den~Nest}},\ }\bibfield  {title} {\bibinfo {title} {Classical simulation
  complexity of extended {Clifford} circuits},\ }\href
  {https://dl.acm.org/doi/10.5555/2638682.2638689} {\bibfield  {journal}
  {\bibinfo  {journal} {Quantum Info. Comput.}\ }\textbf {\bibinfo {volume}
  {14}},\ \bibinfo {pages} {633–648} (\bibinfo {year} {2014})}\BibitemShut
  {NoStop}%
\bibitem [{dat()}]{data_available}%
  \BibitemOpen
  \href@noop {} {}\bibinfo {note} {Code and setup for the simulations is
  available at \url{https://zenodo.org/record/8364056}}\BibitemShut {NoStop}%
\bibitem [{\citenamefont {Walsh}(1963)}]{WalshKSDiscrete}%
  \BibitemOpen
  \bibfield  {author} {\bibinfo {author} {\bibfnamefont {J.~E.}\ \bibnamefont
  {Walsh}},\ }\bibfield  {title} {\bibinfo {title} {Bounded probability
  properties of {Kolmogorov-Smirnov} and similar statistics for discrete
  data},\ }\href {https://doi.org/10.1007/BF02865912} {\bibfield  {journal}
  {\bibinfo  {journal} {Annals of the Institute of Statistical Mathematics}\
  }\textbf {\bibinfo {volume} {15}},\ \bibinfo {pages} {153} (\bibinfo {year}
  {1963})}\BibitemShut {NoStop}%
\bibitem [{\citenamefont {Virtanen}\ \emph {et~al.}(2020)\citenamefont
  {Virtanen}, \citenamefont {Gommers}, \citenamefont {Oliphant}, \citenamefont
  {Haberland}, \citenamefont {Reddy}, \citenamefont {Cournapeau}, \citenamefont
  {Burovski}, \citenamefont {Peterson}, \citenamefont {Weckesser},
  \citenamefont {Bright}, \citenamefont {{van der Walt}}, \citenamefont
  {Brett}, \citenamefont {Wilson}, \citenamefont {Millman}, \citenamefont
  {Mayorov}, \citenamefont {Nelson}, \citenamefont {Jones}, \citenamefont
  {Kern}, \citenamefont {Larson}, \citenamefont {Carey}, \citenamefont {Polat},
  \citenamefont {Feng}, \citenamefont {Moore}, \citenamefont {{VanderPlas}},
  \citenamefont {Laxalde}, \citenamefont {Perktold}, \citenamefont {Cimrman},
  \citenamefont {Henriksen}, \citenamefont {Quintero}, \citenamefont {Harris},
  \citenamefont {Archibald}, \citenamefont {Ribeiro}, \citenamefont
  {Pedregosa}, \citenamefont {{van Mulbregt}},\ and\ \citenamefont {{SciPy 1.0
  Contributors}}}]{ScipyRef}%
  \BibitemOpen
  \bibfield  {author} {\bibinfo {author} {\bibfnamefont {P.}~\bibnamefont
  {Virtanen}}, \bibinfo {author} {\bibfnamefont {R.}~\bibnamefont {Gommers}},
  \bibinfo {author} {\bibfnamefont {T.~E.}\ \bibnamefont {Oliphant}}, \bibinfo
  {author} {\bibfnamefont {M.}~\bibnamefont {Haberland}}, \bibinfo {author}
  {\bibfnamefont {T.}~\bibnamefont {Reddy}}, \bibinfo {author} {\bibfnamefont
  {D.}~\bibnamefont {Cournapeau}}, \bibinfo {author} {\bibfnamefont
  {E.}~\bibnamefont {Burovski}}, \bibinfo {author} {\bibfnamefont
  {P.}~\bibnamefont {Peterson}}, \bibinfo {author} {\bibfnamefont
  {W.}~\bibnamefont {Weckesser}}, \bibinfo {author} {\bibfnamefont
  {J.}~\bibnamefont {Bright}}, \bibinfo {author} {\bibfnamefont {S.~J.}\
  \bibnamefont {{van der Walt}}}, \bibinfo {author} {\bibfnamefont
  {M.}~\bibnamefont {Brett}}, \bibinfo {author} {\bibfnamefont
  {J.}~\bibnamefont {Wilson}}, \bibinfo {author} {\bibfnamefont {K.~J.}\
  \bibnamefont {Millman}}, \bibinfo {author} {\bibfnamefont {N.}~\bibnamefont
  {Mayorov}}, \bibinfo {author} {\bibfnamefont {A.~R.~J.}\ \bibnamefont
  {Nelson}}, \bibinfo {author} {\bibfnamefont {E.}~\bibnamefont {Jones}},
  \bibinfo {author} {\bibfnamefont {R.}~\bibnamefont {Kern}}, \bibinfo {author}
  {\bibfnamefont {E.}~\bibnamefont {Larson}}, \bibinfo {author} {\bibfnamefont
  {C.~J.}\ \bibnamefont {Carey}}, \bibinfo {author} {\bibfnamefont
  {{\.I}.}~\bibnamefont {Polat}}, \bibinfo {author} {\bibfnamefont
  {Y.}~\bibnamefont {Feng}}, \bibinfo {author} {\bibfnamefont {E.~W.}\
  \bibnamefont {Moore}}, \bibinfo {author} {\bibfnamefont {J.}~\bibnamefont
  {{VanderPlas}}}, \bibinfo {author} {\bibfnamefont {D.}~\bibnamefont
  {Laxalde}}, \bibinfo {author} {\bibfnamefont {J.}~\bibnamefont {Perktold}},
  \bibinfo {author} {\bibfnamefont {R.}~\bibnamefont {Cimrman}}, \bibinfo
  {author} {\bibfnamefont {I.}~\bibnamefont {Henriksen}}, \bibinfo {author}
  {\bibfnamefont {E.~A.}\ \bibnamefont {Quintero}}, \bibinfo {author}
  {\bibfnamefont {C.~R.}\ \bibnamefont {Harris}}, \bibinfo {author}
  {\bibfnamefont {A.~M.}\ \bibnamefont {Archibald}}, \bibinfo {author}
  {\bibfnamefont {A.~H.}\ \bibnamefont {Ribeiro}}, \bibinfo {author}
  {\bibfnamefont {F.}~\bibnamefont {Pedregosa}}, \bibinfo {author}
  {\bibfnamefont {P.}~\bibnamefont {{van Mulbregt}}},\ and\ \bibinfo {author}
  {\bibnamefont {{SciPy 1.0 Contributors}}},\ }\bibfield  {title} {\bibinfo
  {title} {{SciPy} 1.0: Fundamental algorithms for scientific computing in
  {Python}},\ }\href {https://doi.org/10.1038/s41592-019-0686-2} {\bibfield
  {journal} {\bibinfo  {journal} {Nature Methods}\ }\textbf {\bibinfo {volume}
  {17}},\ \bibinfo {pages} {261} (\bibinfo {year} {2020})}\BibitemShut
  {NoStop}%
\bibitem [{\citenamefont {Smirnov}(1948)}]{Sm48}%
  \BibitemOpen
  \bibfield  {author} {\bibinfo {author} {\bibfnamefont {N.}~\bibnamefont
  {Smirnov}},\ }\bibfield  {title} {\bibinfo {title} {Table for estimating the
  goodness of fit of empirical distributions},\ }\href
  {http://www.jstor.org/stable/2236278} {\bibfield  {journal} {\bibinfo
  {journal} {Ann. Math. Statist.}\ }\textbf {\bibinfo {volume} {19(2)}},\
  \bibinfo {pages} {279} (\bibinfo {year} {1948})}\BibitemShut {NoStop}%
\bibitem [{\citenamefont {Knuth}(1997)}]{knuth97}%
  \BibitemOpen
  \bibfield  {author} {\bibinfo {author} {\bibfnamefont {D.~E.}\ \bibnamefont
  {Knuth}},\ }\href@noop {} {\emph {\bibinfo {title} {The Art of Computer
  Programming, Volume 2: Seminumerical Algorithms}}}\ (\bibinfo  {publisher}
  {Addison-Wesley},\ \bibinfo {year} {1997})\BibitemShut {NoStop}%
\bibitem [{\citenamefont {Spee}\ \emph {et~al.}(2018)\citenamefont {Spee},
  \citenamefont {Schwaiger}, \citenamefont {Giedke},\ and\ \citenamefont
  {Kraus}}]{SpSc18}%
  \BibitemOpen
  \bibfield  {author} {\bibinfo {author} {\bibfnamefont {C.}~\bibnamefont
  {Spee}}, \bibinfo {author} {\bibfnamefont {K.}~\bibnamefont {Schwaiger}},
  \bibinfo {author} {\bibfnamefont {G.}~\bibnamefont {Giedke}},\ and\ \bibinfo
  {author} {\bibfnamefont {B.}~\bibnamefont {Kraus}},\ }\bibfield  {title}
  {\bibinfo {title} {Mode entanglement of {Gaussian} fermionic states},\ }\href
  {https://doi.org/10.1103/PhysRevA.97.042325} {\bibfield  {journal} {\bibinfo
  {journal} {Phys. Rev. A}\ }\textbf {\bibinfo {volume} {97}},\ \bibinfo
  {pages} {042325} (\bibinfo {year} {2018})}\BibitemShut {NoStop}%
\end{thebibliography}
\end{document}